\documentclass[manuscript]{aastex62}

\usepackage{multirow}
\usepackage[multiple]{footmisc}
\usepackage{gensymb}
\usepackage[para,online,flushleft]{threeparttable}
\usepackage{geometry}
\usepackage{amsmath}

\newcommand{\rgc}{R_{\rm GC}}
\newcommand{\Vr}{V_{R}}
\newcommand{\Vz}{V_{z}}

\newcommand{\Vtot}{V_{\rm tot}}
\newcommand{\kms}{km$\cdot$s$^{-1}$}

\submitjournal{ApJ}

\begin{document}

\title{The Origins of Young Stars in the Direction of the Leading Arm of the Magellanic Stream: Abundances, Kinematics, and Orbits \footnote{Based on observations with the 6.5 m Clay telescope at Las Campanas
Observatory, Chile (program ID: CN2016A-94).}
}

\author{Lan Zhang}
\affiliation{Key Lab of Optical Astronomy, National Astronomical Observatories, CAS, 20A Datun Road, Chaoyang District, 100012 Beijing, China}

\author{Dana I. Casetti-Dinescu}
\affiliation{Department of Physics, Southern Connecticut State University, 501 Crescent St., New Haven, CT 06515, USA}
\affiliation{Astronomical Institute of the Romanian Academy, str. Cutitul de Argint 5, Bucharest, Romania}
\affiliation{Radiology and Biomedical Imaging, Yale School of Medicine, 300 Cedar Street, New Haven 06519, USA}

\author{Christian Moni Bidin}
\affiliation{Instituto de Astronom\'{i}a, Universidad Cat\'{o}lica del Norte, Av. Angomos 0610, Antofagasta, Chile}

\author{R\'{e}ne A. M\'{e}ndez}
\affiliation{Departamento de Astronomia, Universidad de Chile, Casilla 36-D, Santiago, Chile}

\author{Terrence M. Girard}
\affiliation{14 Dunn Rd, Hamden, Connecticut, CT 06518, USA}

\author{Katherine Vieira}
\affiliation{Centro de Investigaciones de Astronomi\'{a}, Apartado Postal 264, M\'{e}rida 5101-A, Venezuela}

\author{Vladimir I. Korchagin}
\affiliation{Institute of Physics, Southern Federal University, Stachki st/194, 344090, Rostov-on-Don, Russia}

\author{William F. van Altena}
\affiliation{Astronomy Department, Yale University, 260 Whitney Ave. , New Haven, CT 06511, USA}

\author{Gang Zhao}
\affiliation{Key Lab of Optical Astronomy, National Astronomical Observatories, CAS, 20A Datun Road, Chaoyang District, 100012 Beijing, China}



\begin{abstract}
We explore the origins of the young B-type stars found by \citet{case14} at the outskirts of the Milky-Way disk in the sky region of Leading Arm of the Magellanic Stream. High-resolution spectroscopic observations made with the MIKE instrument on the Magellan Clay 6.5m telescope for nine stars are added to the previous sample analyzed by \citet{zhang17}. We compile a sample of fifteen young stars with well-determined stellar types, ages, abundances and kinematics. With proper motions from Gaia DR2 we also derive orbits in a realistic Milky-Way potential. We find that our previous radial-velocity selected LA candidates have substantial orbital angular momentum. The substantial amount of rotational component for these stars is in contrast with the near-polar Magellanic orbit, thus rendering these stars unlikely members of the LA. There are four large orbital-energy stars in our sample. The highest orbital-energy one has an age shorter than the time to disk crossing, with a birthplace $z=2.5$~kpc and $R_{\rm GC}\sim 28$~kpc. Therefore, the origin of this star is uncertain. The remaining three stars have disk runaway origin with birthplaces between 12 and 25 kpc from the Galactic center. Also, the most energetic stars are more metal poor ([Mg/H] =$-0.50\pm0.07$) and with larger He scatter ($\sigma_{\rm [He/H]} = 0.72$) than the inner disk ones ([Mg/H] $=0.12\pm0.36$, $\sigma_{\rm [He/H]} = 0.15$). While the former group's abundance is compatible with that of the Large Magellanic Cloud, it could also reflect the metallicity gradient of the MW disk and their runaway status via different runaway mechanisms.

\end{abstract}



%____________________________________________________________________________________
%Section 1: Introduction
%____________________________________________________________________________________
\section{Introduction}
\label{sec:intro}
Recent studies attempting to describe the nearest gas-rich galaxy interaction --- the Magellanic Clouds with each other and the Milky Way (MW) --- are based on a wide range of new and improved observations including proper motions \citep{kall13}, gas content and abundances \citep[e.g.,][]{fox14, fox18, rich18}, and low surface-brightness features obtained with observations with the Dark Energy Camera on the Blanco 4m telescope \citep[e.g.,][]{mac18, nide18} or other facilities \citep{besla16}. Some of these new models \citep[e.g.,][]{hamm15, pard18} while succeeding in some aspects to capture the complex physics of this interaction, they also point to outstanding questions, such as the total gas content of the Magellanic Stream and its relation to present-day gas content in the Clouds \citep[see discussion in][]{pard18}, and the formation of the Leading Arm \citep[LA, e.g.,][]{putm98,nide10,dong16}. Practically, no models have been able to reproduce the morphology of the LA, which has a width of about $60\arcdeg$ and a filamentary structure organized in four different substructures in the sky \citep{venz12, for13}. While most recent models struggle to bring Magellanic stream material as close as $\sim 25$ kpc from the Galactic center (GC) \citep[e.g.,][]{pard18}, evidence from the gas \citep{mccl08, rich18} and from certain young stars at the edge of the MW disk \citep[][hereafter CD14]{case14} suggests that Magellanic Stream material as close as $\sim 20$ kpc from the GC is present. Therefore, any further observational information regarding the LA is beneficial for understanding its formation.

In this contribution we focus on further analyzing the young stars proposed by CD14 to be formed at the edge of the MW's disk from material in the LA as the stream reaches and interacts with the MW's disk \citep[see also][]{mccl08}. Candidate young O/B-type stars for spectroscopic follow-up were drawn from the study by \citet{case12}. CD14 analyzed the radial velocities (RVs) and spectral properties of 42 O/B candidates using intermediate-resolution spectra, and found five young stars with large RVs, compatible with membership to the LA and with a low velocity dispersion of 33  km~s$^{-1}$.

In a subsequent study, \citet[][hereafter Z17]{zhang17} used high-resolution spectroscopy to determine the chemical compositions and kinematics for eight OB-type stars selected from CD14. They identified five B-type stars that have radial velocities compatible with membership to the LA and found that the abundance of these stars --- as reflected by Mg --- is LMC-like, and significantly more metal-poor than that of the stars that kinematically were MW-disk members. 

In this study we add nine more stars from the CD14 list. Specifically, we determine abundances of He, C, N, O, Mg, Si and S using high-resolution spectra obtained with the Magellan Inamori Kyocera Echelle \citep[MIKE;][]{bern03} on the Magellan Clay 6.5m telescope. Together with the stars from Z17 we assemble a well-classified sample of fifteen young stars of which seven have RVs compatible with LA membership. The sample was chosen such that typical disk stars are present in it for comparison purposes. To the spectroscopic data, we add the proper motions from the second data release \citep{gaia_dr2} of \citet{gaia}. Based on this We present a comprehensive orbit and abundance analysis of these stars, as well as a discussion of their origin.

The observations are described in \S~\ref{sec:obs}. The method and results of the measurements of stellar parameters, abundance and distance are described in \S~\ref{sec:meas} and \ref{sec:spec_results}. The 3D kinematics and orbit calculations are presented in \S~\ref{sec:kinematics}. Finally, the possible origins of these stars are discussed in \S~\ref{sec:disc}, with a summary in $\S $~\ref{sec:sum}.

%____________________________________________________________________________________
%Section 2: Data set
%____________________________________________________________________________________
\section{Observations \& Data Reduction}
\label{sec:obs}

\subsection{Target selection}
\label{subsec:star_sele}
Similar to Z17, five stars with heliocentric distances $D < 21$~kpc and radial velocities (RVs) $\gtrsim 90$~km~s$^{-1}$ were selected from the pilot, intermediate-resolution spectral study of CD14. Four stars located in similar regions and with RV $< 90$~km~s$^{-1}$ were also included for the purpose of comparison. Thus, the present study targeted nine stars, distributed above and below the Galactic plane in LA regions LA~I, LA~II, and LA~III \citep{nide08,venz12}. We designate our stars as ``CD14-A**'',  ``CD14-B**'', and ``CD14-C**'' for those at Magellanic coordinates ($\Lambda_M$, $B_M$) $\sim$ ($15^{\circ}$, $-22^{\circ}$), ($\Lambda_M$, $B_M$) $\sim$ ($42^{\circ}$, $-8^{\circ}$), and ($\Lambda_M$, $B_M$) $\sim$ ($52^{\circ}$, $28^{\circ}$), respectively. The detailed spatial distribution of the stars can be seen in Figure~1 of CD14, and we do not repeat it here. We list our stars in Table~\ref{tab:obs_log}, which includes the current designation, the SPM4 identification number, equatorial coordinates, $V$ magnitudes, as well as other observational details. 
 
\subsection{High-resolution spectroscopy and data reduction}
\label{subsec:reduction}
High-resolution spectra of the nine target stars were obtained with the MIKE instrument on the 6.5m Clay telescope in April 2016. The setup gave a resolution of $R \sim 33000$ and $R \sim 29000$ for the blue ($3200 - 5000$~{\AA}) and the red ($4900 - 10000$~{\AA}) sides, respectively. The average seeing and airmass during the observations were 0''.8 and 1.2. Table~\ref{tab:obs_log} summarizes the observational details, exposure times, and the achieved S/Ns of the spectra in the range of 4000 and 5800 {\AA}. 

\begin{figure}
\centering
\includegraphics[width=0.8\textwidth]{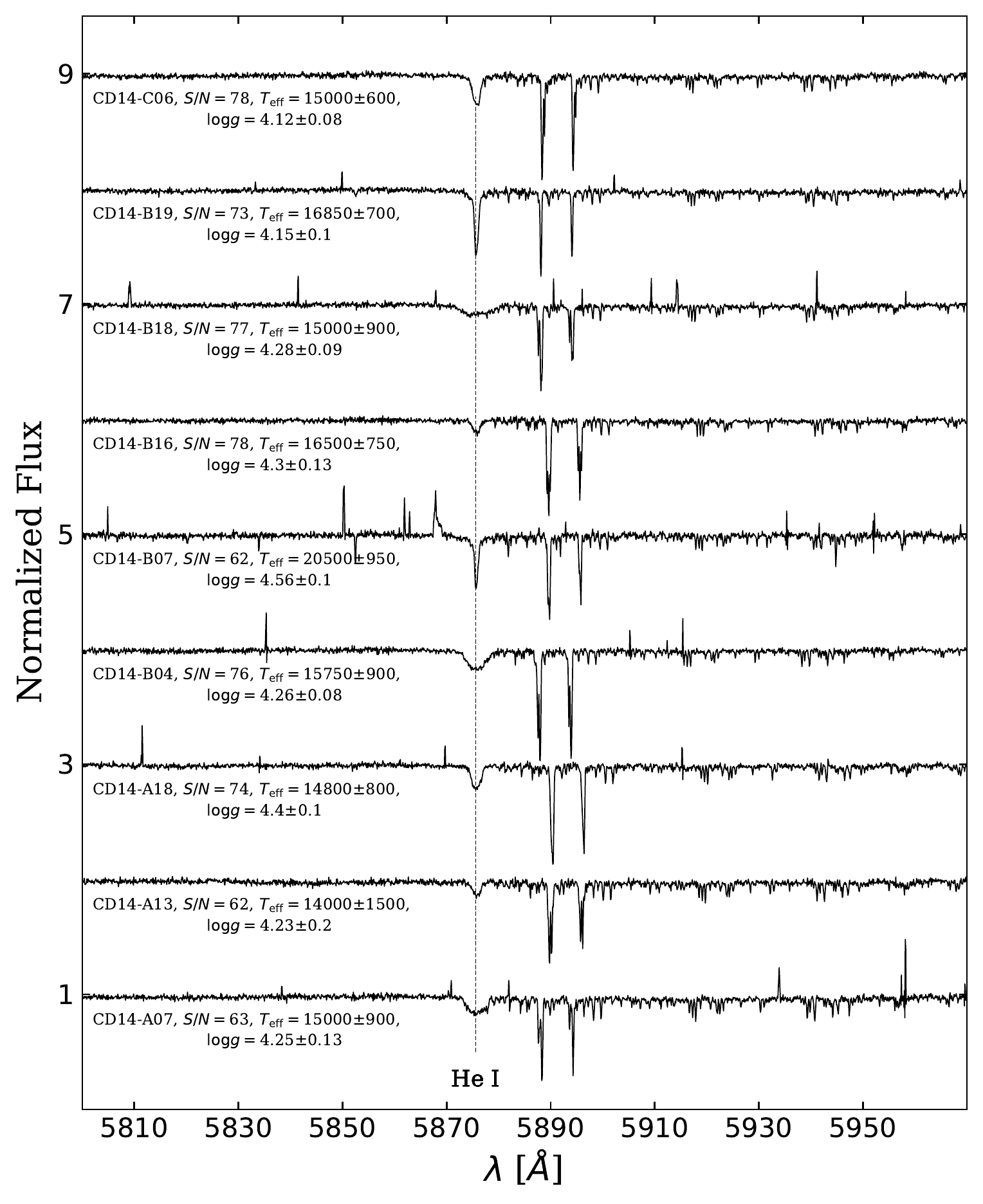}
\caption{Examples of spectra obtained with the Magellan/MIKE for the nine target stars. The vertical dashed line indicates the position of \ion{He}{1} (5875~{\AA}).}
\label{fig:sample}
\end{figure}

The standard Carnegie Python distribution\footnote{\url{http://code.obs.carnegiescience.edu}} \citep{kels00,kels03} routines for MIKE were used for data reduction, including order identification, wavelength calibration, flat-field correction, background subtraction, one-dimensional spectra extraction, and flux calibration. The RVs of our targets were measured  after continuum rectification, by cross-correlation with a synthetic template whose temperature and gravity is similar to those of the targets. The standard algorithm described by \citet{tonr79} was adopted, and the synthetic template was taken from the synthetic spectral library of \citet{muna05}. Two telluric oxygen A bands, the (0,1) transition at 6884~{\AA}, and the (0,0) transition band at 7621~{\AA}, were also adopted to correct the RV zero point, where the molecular data were taken from the HIgh-resolution TRANsmission molecular absorption (HITRAN) database\footnote{\url{https://www.cfa.harvard.edu/hitran/}}. Examples of a portion of the spectra are shown in Figure~\ref{fig:sample} and the obtained heliocentric RVs are listed in Table~\ref{tab:rv_dis_age}.

%____________________________________________________________________________________
%Section 3: Measurements
%____________________________________________________________________________________
\section{Measurements}
\label{sec:meas}
We used the same methodology described in Z17 to measure the stellar parameters, to determine the elemental abundances, and to estimate the corresponding uncertainties. Here we briefly describe the procedures, while more details can be found in Z17 and reference therein.

\subsection{Stellar Atmosphere Parameters}
\label{subsec:sp}
\begin{figure}
\centering
\includegraphics[width=\textwidth]{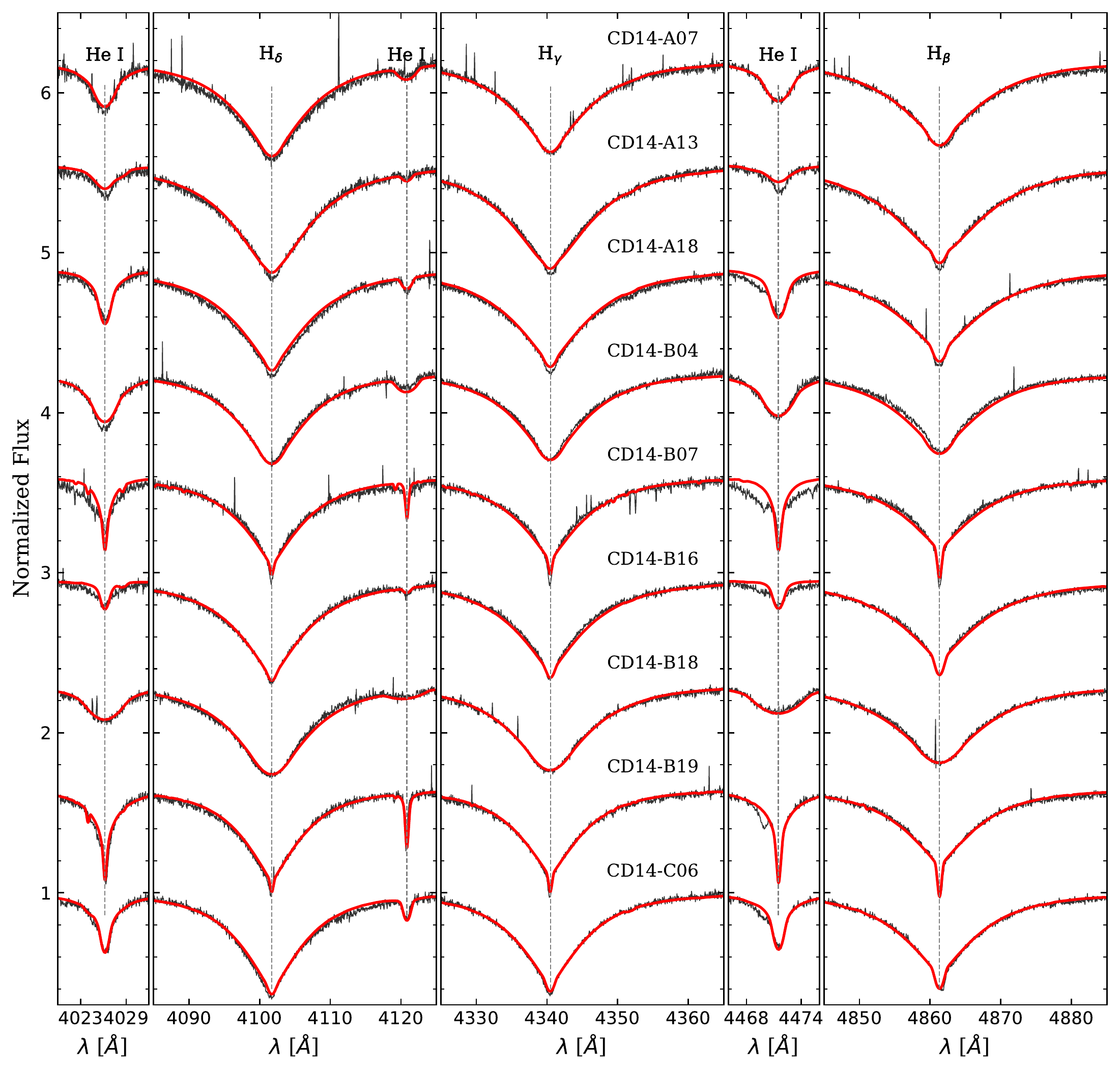}
\caption{Examples of fits to the most prominent spectral features using the best-fit stellar atmosphere parameters (SA I). Gray and red thick lines represent observed and the best synthetic spectra, respectively.}
\label{fig:sp}
\end{figure}

The stellar model atmospheres we employed for our target stars are interpolated from comprehensive grids of metal line-blanketed, non local thermodynamic equilibrium (NLTE), plane-parallel, hydrostatic model atmospheres of B-type stars \citep[][BSTAR2006]{lanz07}, with opacity sampling which is a simple Monte Carlo-like sampling of the superline cross sections and solar abundance. For all our targets, the solar-metallicity ($Z/Z_{\odot}=1$) grid is adopted. However, for CD14-A07, CD14-A18, CD14-B04 and CD14-B18, which show strong helium lines, an additional super-solar metallicity ($Z/Z_{\odot}=2$) grid was also employed for comparison purpose. We have used the spectrum synthesis program SYNSPEC, developed by Ivan Hubeny \& Thierry Lanz\footnote{\url{http://nova.astro.umd.edu/Synspec49/synspec.html}}. The automatic NLTE treatment mode of SYNSPEC is employed, in which the program automatically decides which levels are treated in NLTE, and assigns proper NLTE populations to the lower and upper levels of a given transition. The input model atoms and ions \citep{lanz03b} are provided by TLUSTY\footnote{\url{http://nova.astro.umd.edu/Tlusty2002/tlusty-frames-data.html}}. Balmer series from H$_{\alpha}$ to H$_{10}$ --- except H$_{\epsilon}$ which is blended with \ion{Ca}{1} H line --- and nine \ion{He}{1} lines in ranges of 4000~{\AA} -- 6000~{\AA} are included in the spectrum synthesis procedure for the measurements of stellar parameters. The best fitted stellar parameters are found by a $\chi^2$ test, and the uncertainties estimated from the $\chi^2$ statistics were multiplied by three to obtain the final error, since the errors propagated from the data reduction procedure, such as sky subtraction and the normalization, can not be neglected \citep[][hearafter MB12]{moni12}. Synthetic spectra with the best fitted stellar parameters are shown in Figure~\ref{fig:sp}.

Similar to Z17, we also perform a separate, independent measuring process using different grids of model atmospheres and analysis code (specifically, that in MB12) for the stellar parameters, in order to test the reliability of our stellar-parameter results. The grid of model spectra is computed with Lemke's version\footnote{\url{http://a400.sternwarte.uni-erlangen.de/~ai26/linfit/linfor.html}} of the LINFOR program (developed originally by Holweger, Steffen, and Steenbock at Kiel University), which is based on local thermodynamic equilibrium (LTE) model atmospheres of ATLAS9 \citep{kuru93}. In this analysis, Balmer series from H$_\beta$ to H12, four \ion{He}{1} lines in ranges of 4000~{\AA} -- 5000~{\AA} were fitted simultaneously. For more details on the grids of model atmospheres and analysis code, we refer the reader to MB12 and reference therein. 

\begin{figure}
\centering
\includegraphics[width=\textwidth]{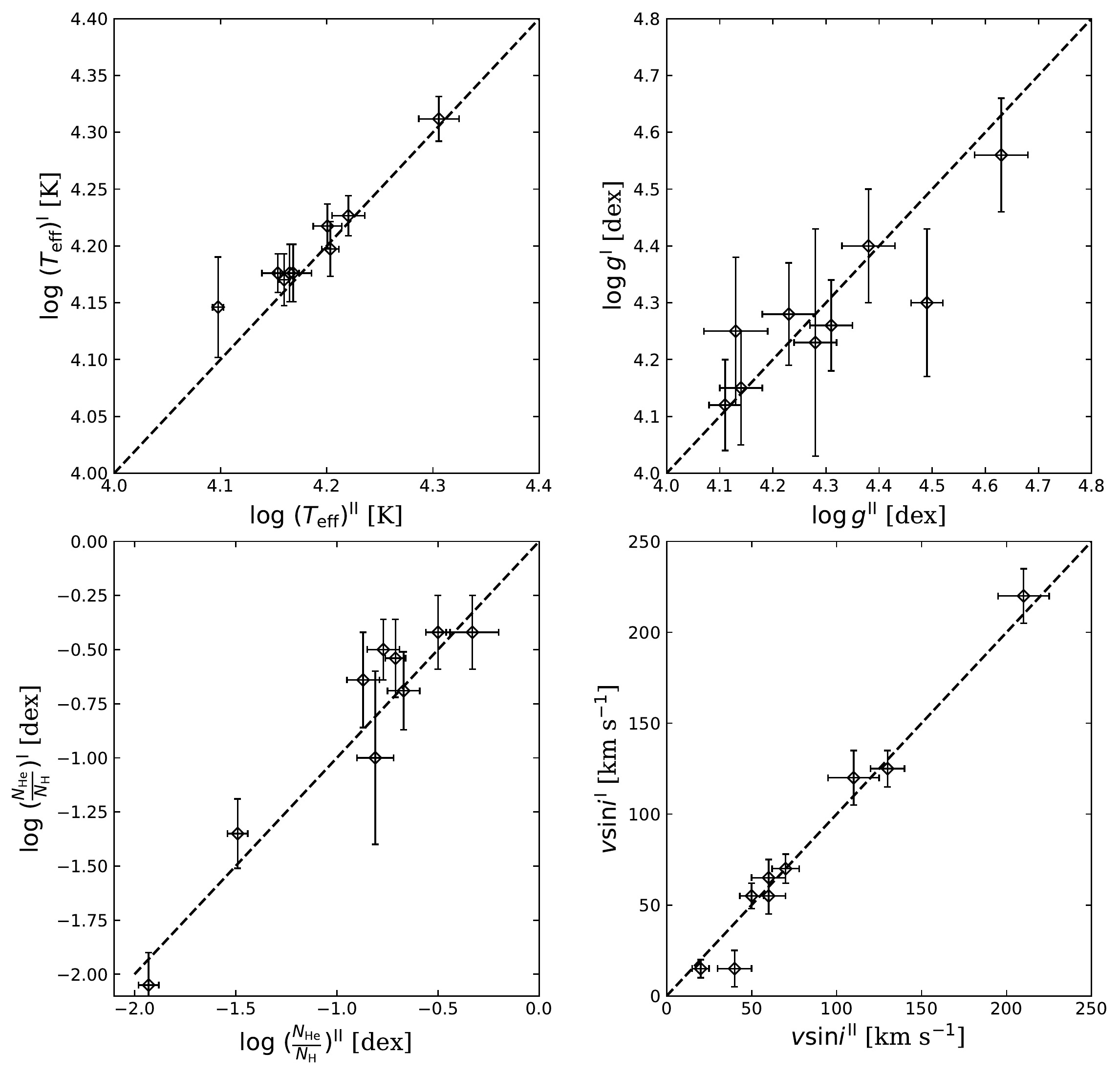}
\caption{Comparison between the two sets of stellar parameters derived from different grids of model atmospheres. The parameters labeled “I” are measured using NLTE grids of TLUSTY (our default method), while those labeled “II” are measured with the separate, independent grid (LTE) and analysis code of LINFOR (as in MB12).}
\label{fig:sp_com}
\end{figure}

Comparisons of the atmospheric parameters between the two independent analyses are also presented in Table~\ref{tab:ste_p} and Figure~\ref{fig:sp_com}. For convenience, we label our default spectral analysis process described in the previous subsection and the separate one {\it \'{a} la} MB12 as ``SA~I'' and ``SA~II'', respectively. The two sets of parameter results reveal an overal good agreement, apart from one problematic case -- CD14-A13. The SA~II temperature of CD14-A13 is lower than the cooler end of the BSTARS grid at 15000~K. Although model atmospheres could be extrapolated from the grid, this is only

valid when when $T_{\rm eff}$ and $\log g$ values are not far from the limits of the grid, that is, small compared to the grid steps. Therefore, the parameter values of SA~I are the best choice given the possible grid ranges of BSTARS. With the exclusion of CD14-A13, the mean difference is a mere 330~K in $T_{\rm eff}$ and $-0.015$~dex in $\log g$. This offset in $T_{\rm eff}$ seems systematic, although negligible compared to errors. Rotational velocities are in good agreement (mean difference of 2~km~s$^{-1}$). For consistency, we will adopt SA~I results for the subsequent abundance analysis.

\subsection{Abundances and uncertainties}
\label{subsec:abundance}
Using the stellar parameters obtained with the stellar model atmospheres and synthesis code described in the previous subsection, we proceed to determine the elemental abundances of Carbon, Nitrogen, Oxygen, Neon, Magnesium, Silicon, and Sulfur by fitting observed lines with synthetic spectra. We only determine the iron abundance for CD14-B07 and CD14-B19 as there are no clear, unambiguous iron lines that can be detected for the other sample stars. For each element, synthetic spectra are produced using the stellar parameters measured with the SA~I procedure and with various abundance values. These synthetic spectra are then compared with the observed ones. A $\chi^2$ test was employed to establish the best fit for detected absorption lines. The mean abundance of each element is derived from all lines that can be clearly detected. For stars in which feature lines for certain elements cannot be detected clearly, the feature that is at the position of the theoretical absorption line was fitted, and the maximum value for this element that could fit the spectrum is considered as the upper limit for its abundance. During the procedure, the solar composition from \citet{grev98} is adopted. The atomic line data of elements are the same as Z17, that is, are mainly selected from \citet[][K95]{kuru95} and the NIST Atomic Spectra Database\footnote{\url{http://www.nist.gov/pml/data/asd.cfm}}. 

\begin{figure}
\centering
\includegraphics[scale=0.85]{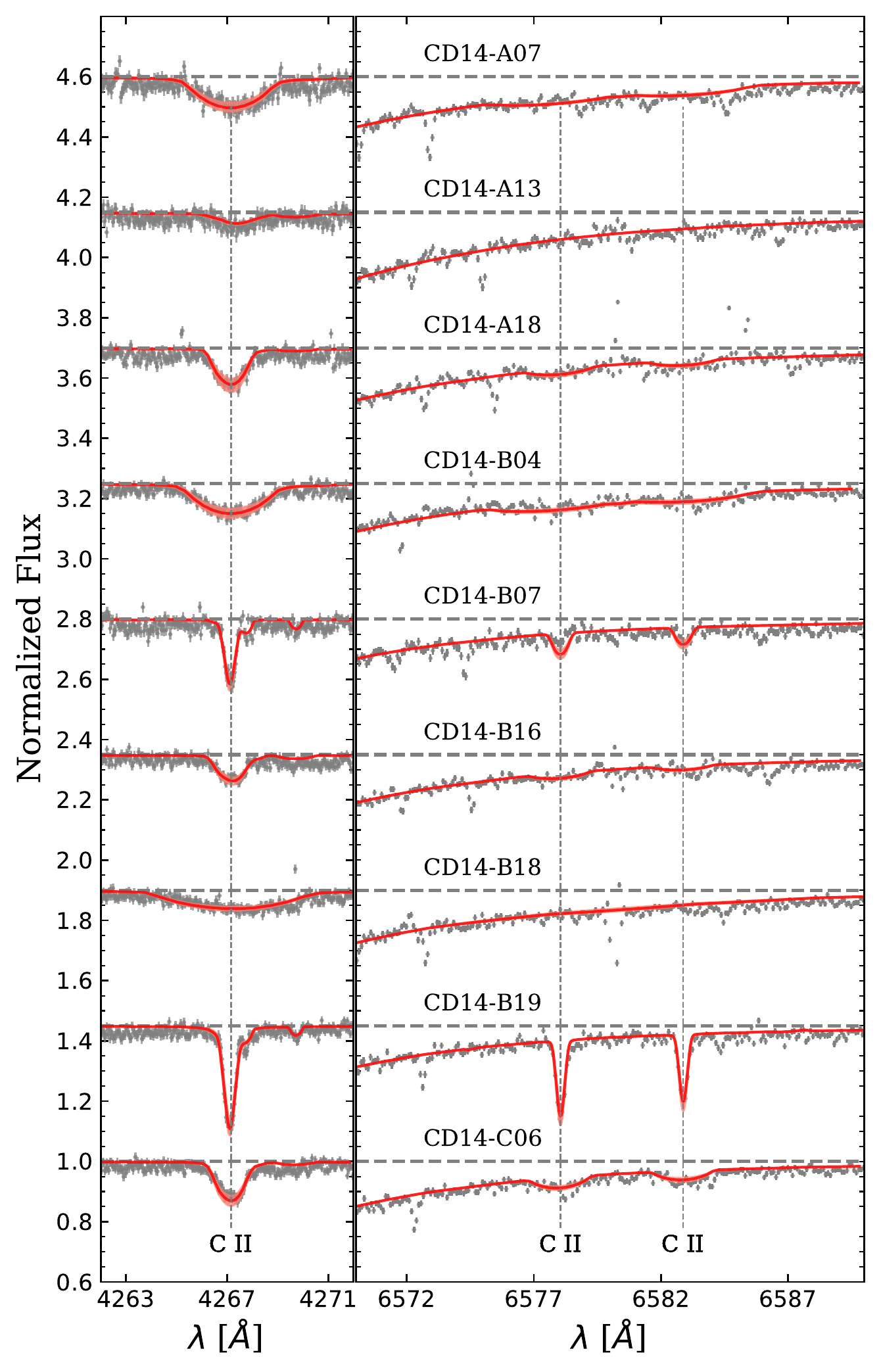}
\caption{Examples of fits of observed spectra (dark dots) with synthetic spectra (thick lines) for the \ion{C}{2} lines of the target stars.}
\label{fig:abun_c}
\end{figure}

\begin{figure}
\centering
\includegraphics[width=\textwidth]{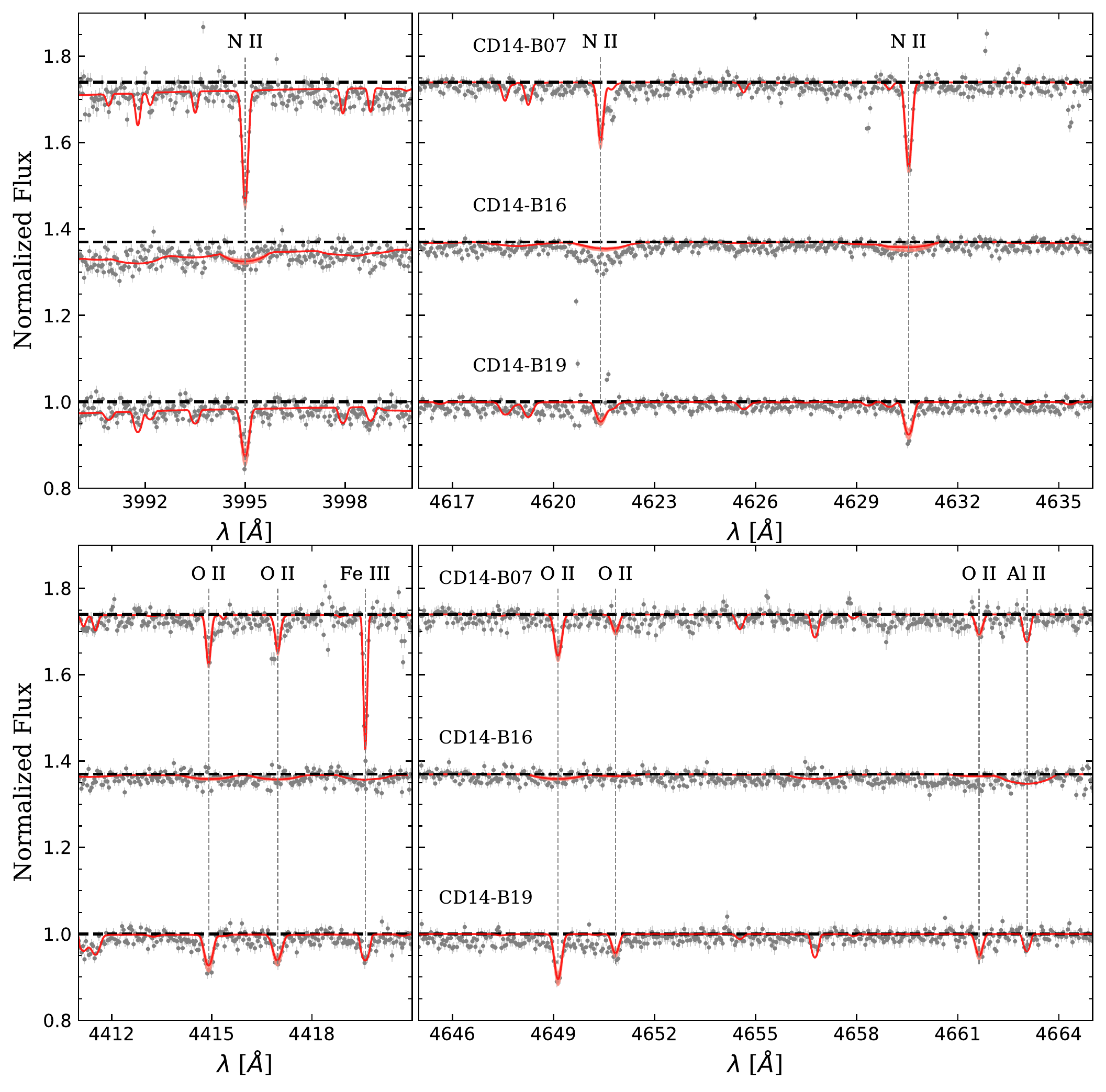}
\caption{Similar to Figure~\ref{fig:abun_c}, but for elements N and O. Only stars for which \ion{O}{2} and \ion{N}{2} lines could be detected are shown.}
\label{fig:abun_no}
\end{figure}

\begin{figure}
\centering
\includegraphics[width=\textwidth]{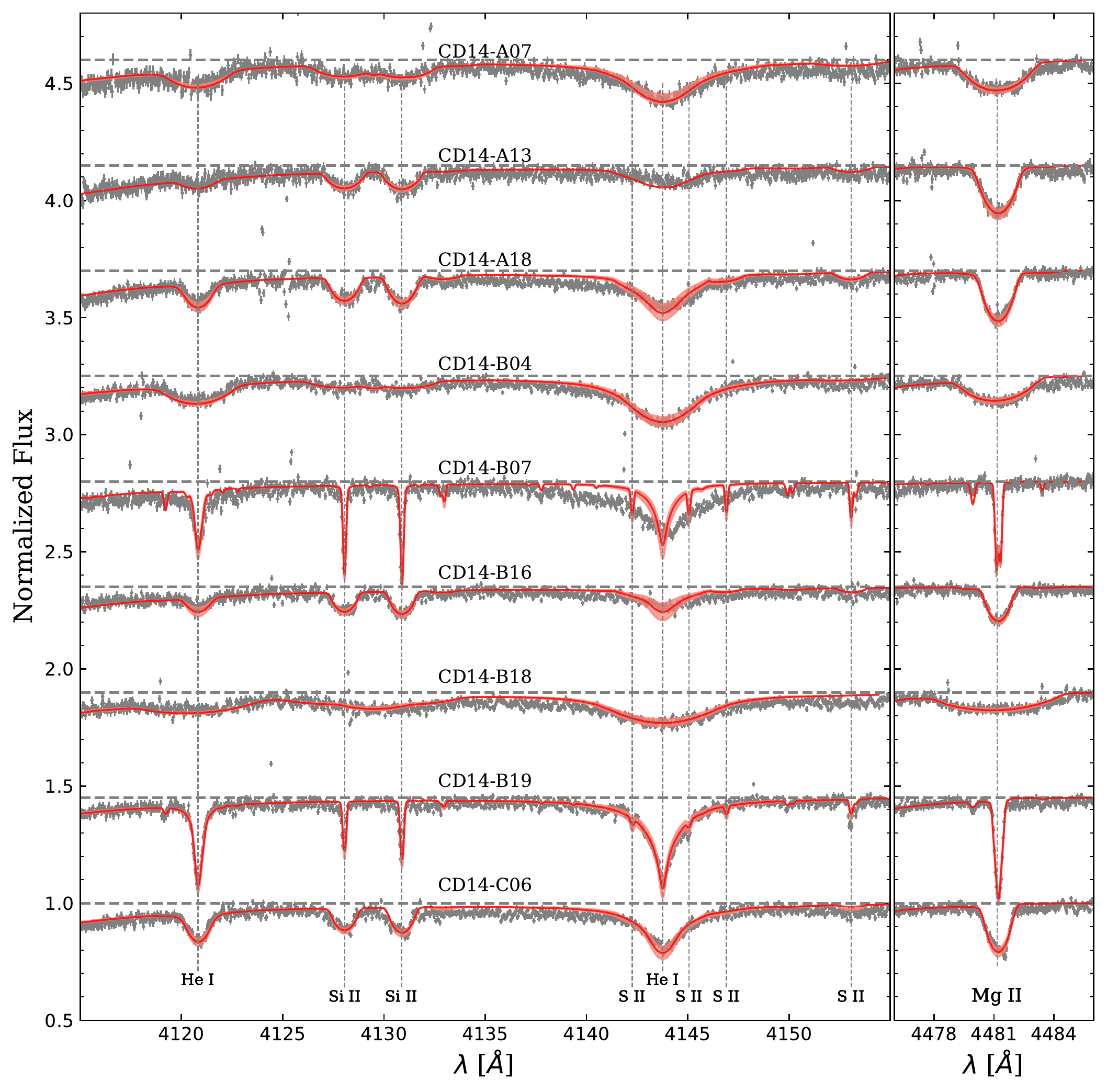}
\caption{Similar to Figure~\ref{fig:abun_c}, but for elements He, Mg, Si, and S.}
\label{fig:abun_mgsis}
\end{figure}

The overall abundance uncertainty is estimated by summing in quadrature the uncertainties from the following sources: 
\begin{itemize}
\item[1)] Uncertainties from the observational error and the fitting procedure. 
\item[2)] Uncertainties from atomic data. In the present study, an uncertainty of 10\% in $\log gf$ was also adopted to explore its effect on the abundance, which results in an error of 0.02 dex on average.
\item[3)] Errors in the continuum rectification. In the worst case, the error in continuum rectification was estimated to be 5\%, which results in a change of the abundance of up to 0.05~dex.
\item[4)] Uncertainties propagated from errors of the stellar parameters. 
\item[5)] The scatter in the abundance determinations from different lines gives another estimate of the uncertainty. 
\end{itemize}
More details can be found in Z17. Synthetic spectra with error ranges that are generated for the best-fit parameters of sample stars are shown in Figures~\ref{fig:abun_c} $-$~\ref{fig:abun_mgsis}, using representative \ion{C}{2}, \ion{N}{2},  \ion{O}{2},  \ion{Mg}{2} 4481 {\AA}, \ion{Si}{2}, and \ion{S}{2} as examples.

\subsection{Distances and ages}
\label{subsec:dis_age}
Gaia DR2 provides parallaxes for $\sim 1.7$ billion of stars in the Milky Way. However, the relative errors are larger than 25\% for stars more distant than 4~kpc from the Sun. For this reason, we instead adopt in the following spectroscopic distances for our sample of stars stars.

The present sample is contaminated by horizontal-brach (HB) stars (see \S~\ref{subsec:classification}). For true main-sequence (MS) stars of the sample, we determine their age via comparison of their position in the temperature-gravity plane with PARSEC isochrones \citep{bres12}. The uncertainty in this estimate is given by the error box associated with the stellar parameters. This is shown in Figure~\ref{fig:sp_track}. The isochrones also return an estimate of the absolute magnitude $M_V$ of each star. For the HB stars, we adopt the HB evolution tracks (PARSEC) to estimate the masses and luminosities, and possibly, the evolution time after post-helium-flash. 

Finally, we calculate the true distance moduli $(m-M)_0$ from the apparent $V$ magnitudes of the SPM4 catalog, de-reddened with the \citet{schlegel98} maps as corrected by \citet{boni00}, and assuming a standard reddening law with $R = \frac{A_V}{E(B-V)}=3.1$. We have verified that the results are not affected within errors by the use of the \citet{schlafly11} calibration, and that the distance moduli obtained using the 2MASS $J$ and $K$ magnitudes are consistent with those obtained in $V$, with no systematics.

%____________________________________________________________________________________
%Section 4: Results I: spectroscopic results
%____________________________________________________________________________________
\section{Spectroscopic Results}
\label{sec:spec_results}

\subsection{Abundances}
\label{subsec:abun}
The measured stellar parameters and abundances are summarized in Tables~2 and \ref{tab:abun_results_a} and shown in Figure~\ref{fig:abun_res}. For stars CD14-A07, CD14-A18, CD14-B04, and CD14-B18, whose detected abundance patterns are all super-solar, the abundance values derived by the metal-rich model atmospheres ($Z/Z_{\odot} = 2$) are also shown (see yellow symbols in Figure~\ref{fig:abun_res}). Abundances derived using the metal-rich model atmosphere appear slightly, but systematically, lower than those obtained using the solar metallicity model atmosphere; however, values are consistent within their $1-\sigma$ errors. For a better analysis, in the following discussion we will adopt the results from metal-rich model atmospheres for these four stars. Except for CD14-B07 and CD14-B16, He and C abundances of the stars are super-solar or near-solar, within uncertainties. Because of rapid $v\sin i$, \ion{S}{2} lines combined with \ion{He}{1} 4143~{\AA} can not be separated clearly for most of the targets. Only an upper limit of S abundance can be estimated by fitting line wings of \ion{He}{1} 4143~{\AA}. We proceed now to discuss individual stellar results.

\subsubsection{CD14-A07, CD14-A18, CD14-B04, and CD14-B18}
\label{subsubsec:a07}
Among the four metal-rich stars, CD14-A07, CD14-B04, and CD14-B18 have large rotational velocities. Weak lines cannot be detected due to the broadening effects of rotation. Only strong lines such as \ion{He}{1} 5587~{\AA}, \ion{Mg}{2} 4487~{\AA} and \ion{C}{2} 4266~{\AA} can be detected and used in the abundance determination. The measurement of stellar parameters and of elemental abundances is made more difficult by having so few spectral features. Therefore, we do not have sufficient information for a reliable abundance determination for these three stars. 
Also, during the line fitting, rotational broadening implies that the line profile is insensitive to the variance of the abundance, which leads relatively large uncertainties of the abundances for the three stars.

Although no usable iron lines can be detected for [Fe/H] measurement, we assume their [Fe/H] is about 0.3~dex because the model-atmospheres used for these four stars is $Z/Z_{\odot}=2$, while [Mg/Fe], [Si/Fe], and [S/Fe] show no obvious enhancements.
 
\subsubsection{CD14-B07 and CD14-B19}
\label{subsubsec:b07}
Most of the absorption lines used in our chemical analysis are detected for these two stars because they are not sensibly affected by the broadening effects of rotation. Strong \ion{N}{2} 3995~{\AA} can be detected in the spectra of CD14-B07 and CD14-B19 implying a significant nitrogen enhancement. The abundance of Mg for the two stars is sub-solar, while Si and S abundances are super-solar and sub-solar for CD14-B07 and CD14-B19, respectively. \ion{Fe}{3} line at 4421~{\AA} can be detected for the two stars, from which we estimated their [\ion{Fe}{3}/H] is about $0.42\pm0.12$~dex and $0.05\pm0.18$~dex, respectively. Besides these lines, one phosphorus line \ion{P}{2} 4588~{\AA} can be detected in the spectra of CD14-B07, yielding an abundance of [\ion{P}{2}/H]$ = 1.23\pm0.10$~dex.

\subsubsection{CD14-A13, CD14-B16, CD14-C06}
\label{subsubsec:a13}
There are no He and C enhancements in CD14-A13. The abundance uncertainties of the star are relatively large because of the large error bars propagated from the stellar parameters. CD14-B16 appears depleted in He and C, and its Mg abundance is similar to the average value of the LMC. Similar to the four metal-rich stars mentioned above, CD14-C06 shows He and C enhancements. The remaining of its abundance values are slightly sub- or near solar, within the uncertainties.

\subsection{Stellar classification}
\label{subsec:classification}
\begin{figure}
\centering
\includegraphics[width=\textwidth]{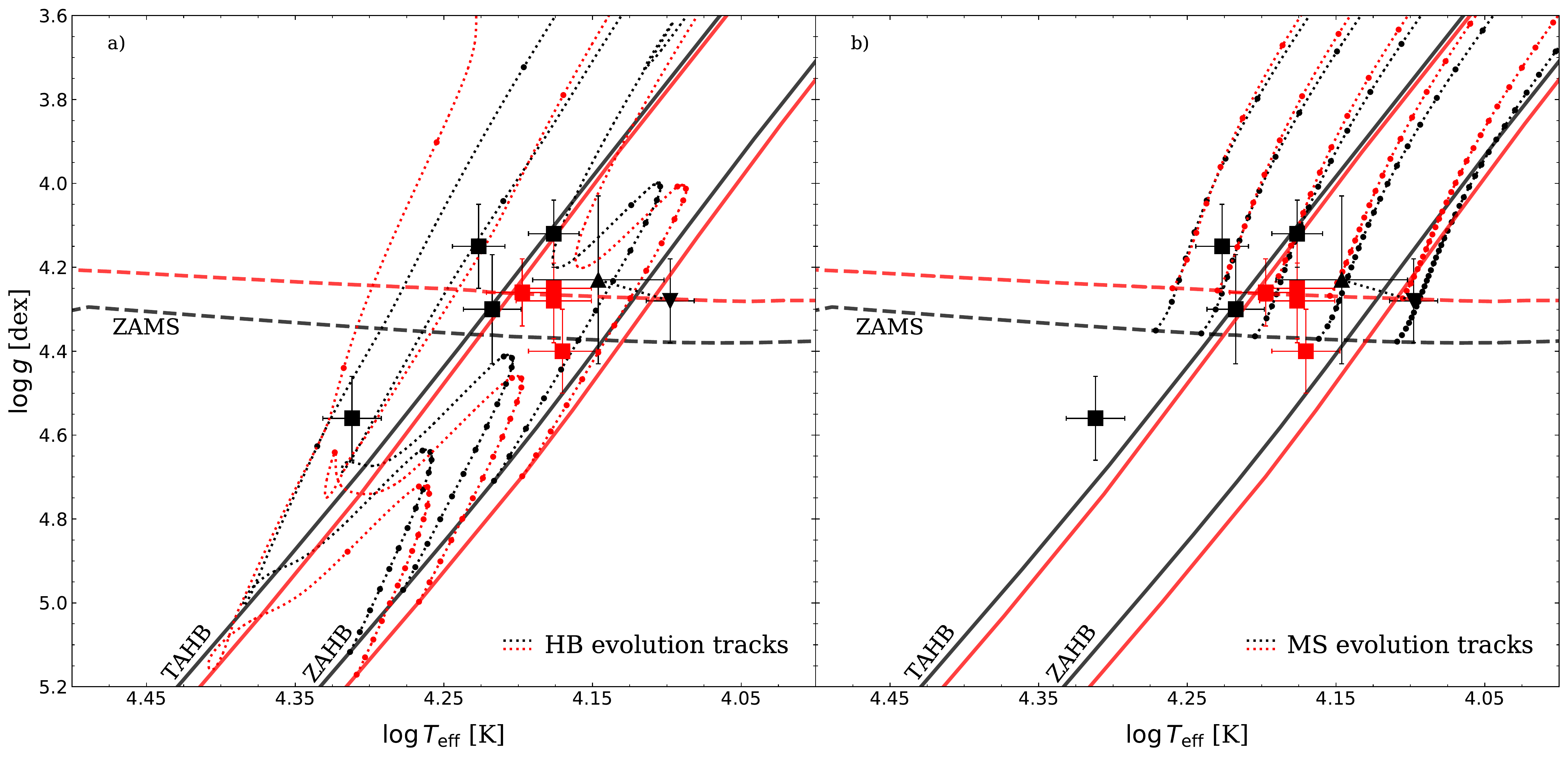}
\caption{The location of target stars with respect to PARSEC evolution tracks \citep{bres12, bres13, chen14} in the temperature-gravity plane. Black and red symbols/lines represent $Z/Z_{\odot}=1$ and $Z/Z_{\odot}=2$, respectively. The zero-age horizontal branch (ZAHB) and terminal-age horizontal branch (TAHB) tracks are shown in solid lines, while zero-age main sequence (ZAMS) are represented by dashed lines. The upward and downward triangles connected by a dotted line represent the location of CD14-A13 with $T_{\rm eff}$ and $\log g$ from SA~I and SA~II, respectively. a) HB evolution tracks are over plotted, corresponding to models with $M = 0.490~M_{\odot},\, 0.495~M_{\odot},$ and $0.505~M_{\odot}$ for $Z/Z_{\odot}=1$, and $M = 0.475~M_{\odot},\, 0.480~M_{\odot},$ and $0.490~M_{\odot}$ for $Z/Z_{\odot}=2$ (bottom left to upper right). Nodes on the tracks indicate the evolution times. The time interval is 10~Myr. b) Similar to a), but MS evolution tracks are over plotted, corresponding to models with $M = 3.0~M_{\odot},\, 3.6~M_{\odot},\, 4.2~M_{\odot},\, 4.8~M_{\odot},$ and $5.4~M_{\odot}$ for $Z/Z_{\odot}=1$, and $M = 3.4~M_{\odot},\, 4.0~M_{\odot},\, 4.6~M_{\odot},\, 5.2~M_{\odot},$ and $5.8~M_{\odot}$ for $Z/Z_{\odot}=2$ (right to left).}
\label{fig:hb_check}
\end{figure}

\begin{figure}
\centering
\includegraphics[width=\textwidth]{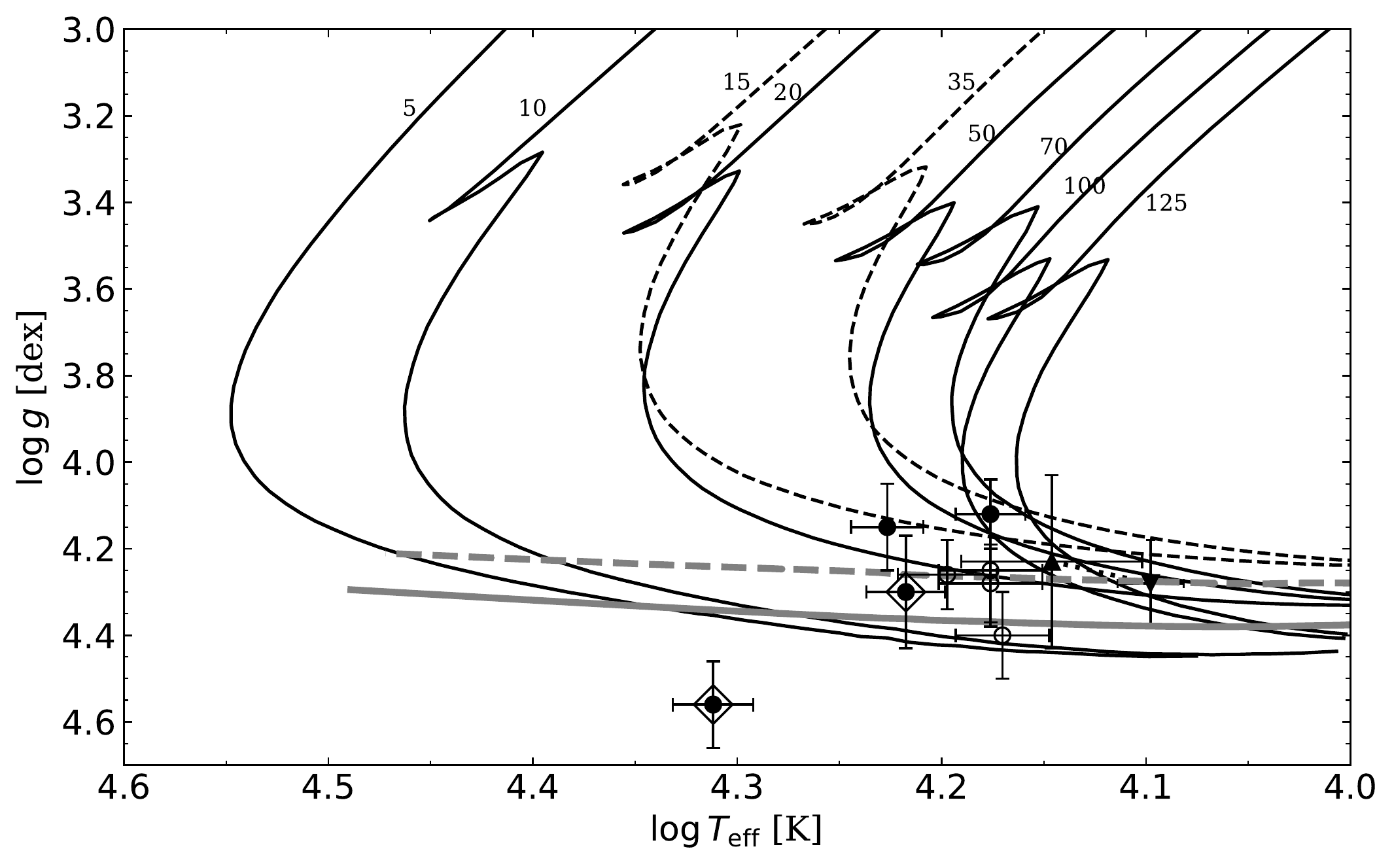}
\caption{The location of target stars with respect to PARSEC isochrones with solar metallicity (black solid lines) and super-solar metallicity (black dashed lines) \citep{bres12}. Ages (in Myr) of each isochrone are labeled. ZAMS with $Z=0.014$ (thick grey solid line) and $Z=0.03$ (thick grey dashed line) are also shown. Open and filled symbols represent stars with solar and super-solar metallicity , respectively. Two HB stars are marked by surrounding diamonds. The upward and downward triangles connected by a dotted line represent the location of CD14-A13 with $T_{\rm eff}$ and $\log g$ from SA~I and SA~II, respectively.}
\label{fig:sp_track}
\end{figure}

\begin{figure}
\centering
\includegraphics[width=\textwidth]{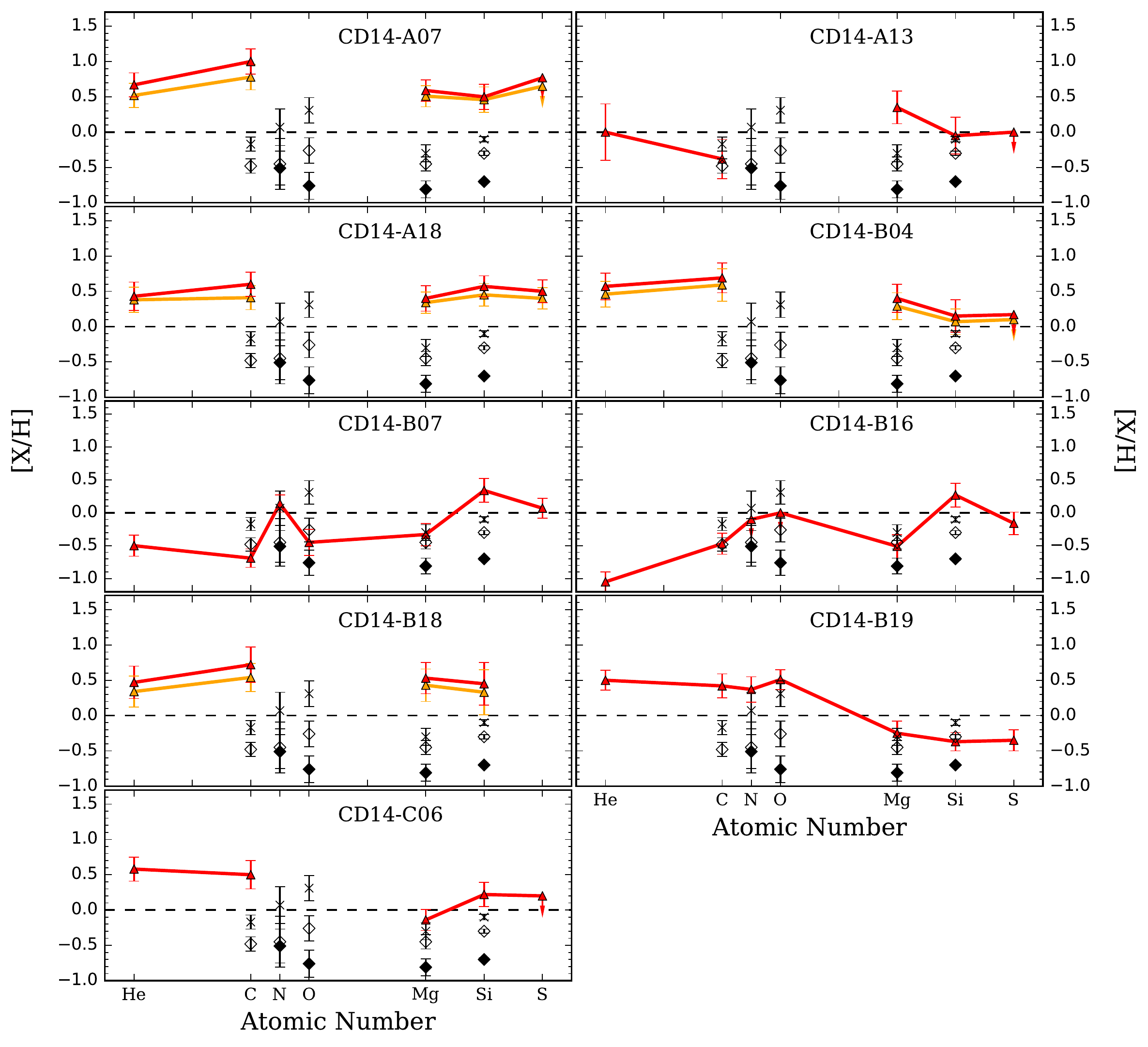}
\caption{Abundance results derived using the stellar parameters from SA I. Red symbols show our results. Orange symbols are results derived from the super-solar metallicity grid of model atmospheres (see text). Downward arrows indicate that only upper limits of the abundances were determined. Weighted average abundances of N, O, Mg, and Si of B-type stars in three clusters are shown as references. Specifically, NGC 4755 (in the MW), NGC 2004 (in the LMC), and  NGC 330 (in the SMC) from \citet{trun07} are represented with crosses, open diamonds, and filled diamonds, respectively. The C abundances of NGC2004-D15 (in the LMC, open diamonds) and HR 3468 (in the solar neighbourhood, crosses) from \citet{przy08b} also are shown for reference.}
\label{fig:abun_res}
\end{figure}

The $\log g$ values of the present targets are larger than 4.0~dex in the $T_{\rm eff}$ range of 15000~K -- 21000~K, in which hot HB and MS B stars occupy overlapping regions of the $T_{\rm eff} - \log g$ diagram. This can be seen in Figure~\ref{fig:hb_check}, where the targets are plotted in the $T_{\rm eff} - \log g$ plane along with HB and MS evolutionary tracks \citep[PARSEC,][]{bres12, bres13, chen14}. In this figure, the stars are separated into two groups according to their metallicities, that is,  CD14-A07, CD14-A18, CD14-B04, and CD14-B18 belong to $Z/Z_{\odot} =2$ (red symbols) while the remaining five stars belong to $Z/Z_{\odot} =1$ (black symbols). To classify each of these stars, abundance and $v\sin i$ info are also considered. A star-by-star analysis reveals that
\begin{enumerate}
\item {\it CD14-A13, CD14-B19, and CD14-C06} -- The three stars lie above the  ZAMS of solar metallicity. Except for CD14-B19, the other two stars are found between the canonical zero-age and terminal-age HB (ZAHB and TAHB, respectively), where post-helium-flash stars spend 99\% of their helium-burning lifetime. MB12 found that the He abundance of HB stars strongly increases with $T_{\rm eff}$ in the range $15000~{\rm K} < T_{\rm eff} < 20000~{\rm K}$, and reaches the minimum value of $\log \frac{N_{\rm He}}{N_{\rm H}} \sim -2.5$ around 15000~K because of the maximum efficiency of diffusion at this temperature. There might however be an offset of $-0.5$~dex for field HB stars compared to the cluster measurements \citep{moni09}. The He abundance of the three stars are all super-solar, which suggests that they are not HB after all. We then check the evolutionary state of CD14-B19. If CD14-B19 were an HB star evolving off the TAHB phase, as shown in Figure~\ref{fig:hb_check}, its He abundance would show depletion to some degree. However, its super -solar [He/H] value ($0.50\pm0.12$) indicates otherwise. Also, from Figure~\ref{fig:hb_check}, one can see that CD14-A13 and CD14-C06 will pass through an evolution phase from $T_{\rm eff}=15000~{\rm K}$, $\log g = 4.3$ to $T_{\rm eff}=20000~{\rm K}$, $\log g = 4.0$ in $\sim$10~Myr if they are HB stars, while the time interval is $\sim$ 100~Myr if they are MS stars. Therefore it is more probable to observe these two stars if they are in their MS evolutionary phase. In addition, both CD14-A13 and CD14-C06 show rapid rotation ($v\sin i > 50$~\kms), which are seldom found in evolved objects \citep[ Figure 8 of ][]{behr03}. We conclude that these three stars are in the MS phase.

\item {\it CD14-B07} -- The locus of the star is far below the ZAMS, and above the region between ZAHB and TAHB. Considering the abundances we see that: 1) both of its magnesium and helium are depleted; 2) iron abundance is enhanced; 3) phosphorus is strongly enhanced. These features are found in hot HB stars in globular clusters as well as in field stars \citep[e.g.,][]{behr99,behr03,moeh99} because of chemical diffusion. That is, iron abundance may be enhanced, while magnesium and helium would show depletion because of radiative levitation of most metal species, and gravitational settling of helium \citep{behr99}. On the contrary, MS stars are usually not affected by atmospheric diffusion. Combining these results with the slow rotation, we classify CD14-B07 as a star with 0.49~$M_{\odot}$ that has just evolved past the post-helium-flash phase.

\item {\it CD14-B16} -- This star lies slightly above the solar matellicity ZAMS. From Figure~\ref{fig:hb_check}, it can be seen that the time intervals CD14-B16 spends in HB and MS tracks are comparable. Although the $v \sin i$ of the star is $55\pm10$~\kms --- which is larger than the upper limits ($35$~\kms) of field HB stars given by \citet{behr03} --- the value is still acceptable for an HB star if a $2-\sigma$ uncertainty is considered. Moreover, its low He abundance ([He/H]$=-1.05\pm0.15$) implies that it could be a post-helium-flash star evolving off the HB. In addition, its abundance patterns are similar to those of CD14-B07 (see Figure.~\ref{fig:abun_res}). The star is therefore marked as an HB object.

\item {\it  CD14-A07, CD14-A18, CD14-B04, and CD14-B18} -- These four metal-rich stars are all in the post-helium-flash region. Although CD14-A18 lies slightly below the metal-rich ZAMS, its $v\sin i $ of $65\pm10$~\kms and super-solar abundance patterns suggest it is not an HB star. Moreover,  if it is actually in the MS phase, from \S~\ref{subsec:dis_age}, it is a metal-rich star located at $\rgc = 6.26~$kpc, which agrees well with the Galactic radial metallicity gradient \citep[e.g.,][]{magr09,silv16}. Therefore, we mark this target as a MS star. The other three stars, CD14-A07, CD14-B04, and CD14-B18 lie on the metal-rich ZAMS. They all rotate very fast with $v\sin i > 100$~\kms, and are therefore very young MS stars. 
\end{enumerate}

In summary, we find that the sample is contaminated by two HB objects, CD14-B07 and CD14-B16. CD14-A18 is a MS star because of its fast $v\sin i$ and super-solar He abundance values, although it lies below the ZAMS models. Therefore, the seven objects in the MS evolutionary phase are further considered in the following discussion.

\subsection{Radial velocities}
\label{subsec:rv_res}
The results of our radial velocity measurements are given in Table~\ref{tab:rv_dis_age}. The error bar of CD14-B18 is relatively large because its absorption lines are broaden by large $v\sin i$ values. A comparison between our RV measurements and the first-epoch values of CD14 is shown in the same table. The two sets of results exhibit good agreement. The RV differences will be taken as ours$-$CD14 hereafter, with errors being the quadratic sum of the uncertainties of the two measurements. All stars show variations lower than the average $0.6-\sigma$ error, and small compared to the huge variations typical of massive B-type binary stars. The only exception is star CD14-B19, whose variation of $\Delta RV=15$~km~s$^{-1}$ is significant at the $2\sigma$ level. 

From \S~\ref{subsec:classification}, it can be seen that seven stars are classified as B-type MS stars. It is known that the fraction of close binaries among B-type stars is greater than 50\% \citep[e.g.,][]{kouw07}. To check the fraction of binary detections given our two-epoch observations and a detection threshold of $\Delta RV=20$~km~s$^{-1}$, we adopt the same model as the one in Z17 (and reference therein), and find that the probability of our null detection out of seven stars is 14\% (8.7\%) if the underlying binary fraction is 50\% (60\%). We did not find any binaries in the sample of Z17. If the current and the Z17 samples are combined, the fraction of null detection turns out to be 2.0\%(0.8\%).

Considering that the $\Delta RV$ value of CD14-B19 is close to the detection threshold adopted here, and the rather low null detection fraction obtained in this calculation, it is possible that CD14-B19 is a binary, and the adopted threshold of $\Delta RV=20$~km~s$^{-1}$ a rather conservative value.

\subsection{Distances and ages}
\label{subsec:dis_age_res}
The results of distance and age estimates are also presented in Table~\ref{tab:rv_dis_age} and Figure~\ref{fig:sp_track}. We used super-solar metallicity isochrones with $Z/Z_{\odot}=2$ for CD14-A07, CD14-B04, and CD14-B18. The results show that the three targets are extremely young stars ($<$ 13 Myr) located at heliocentric distance $d = 4$ to $d = 8$~kpc. Star CD14-A18 would also require the same treatment, but its location in the $T_{\rm eff} - \log g$ plane is slightly below the the metal-rich zero-age MS (ZAMS) model \citep{bres12, bres13} even if the $1-\sigma$ error of $\log g$ is considered. Assuming it is a MS star, its theoretical magnitude is determined at the temperature from the metal-rich ZAMS. This results in a heliocentric distance of 8.5~kpc for CD14-A18.

%____________________________________________________________________________________
%Section 5: Results II: orbits
%____________________________________________________________________________________
\section{3D Kinematics and Orbit Determination}
\label{sec:kinematics}
\subsection{Proper motions}
\label{subsec:pms}
With our measured radial velocities and spectroscopic distances, and with proper motions extracted from Gaia DR2 catalog it is possible to determine complete 3D kinematic information. We do so for the stars in the present study and in Z17. One distant star from CD14, namely CD14-A19, is also included in the following calculation, although no high-resolution spectroscopic information is available. The proper motions and parallaxes from the Gaia DR2 catalog are listed in Table~\ref{tab:pms}: there are nine stars from the current study, eight from Z17 and one from CD14. However, for the velocity and orbit analysis, we will discard stars CD14-B07, CD14-B16 from this study, since they are HB stars and star CD14-A08 from Z17 since it has an ambiguous classification. We are therefore left with fifteen stars for the kinematic and orbit analysis.

\subsection{Velocity components in the Galactic cylindrical coordinate system}
\label{subsec:3d_vel}
We first calculate the rectangular velocity components with respect to the Galactic center (GC) by adopting a value of 238~\kms \citep{scho12} for the local standard of rest ($V_{\rm LSR}$) and a solar motion of (+9.58, +10.52, +7.01)~\kms \citep{tian15} in ($\mathit{U}$, $\mathit{V}$, $\mathit{W}$), which are defined in a right-handed Galactic system with $\mathit{U}$ pointing toward the GC, $\mathit{V}$ in the direction of rotation, and $\mathit{W}$ toward the north Galactic pole \citep{dehn98}. The kinematic parameters including distance and velocities are then transformed to the Galactic cylindrical coordinate system:
\begin{equation}
\begin{aligned}
       & \phi = \arctan \left[\frac{D\cos(b)\sin(l)}{R_{\sun} - D\cos(b)\cos(l)} \right]\\
       & \vec{v}_{\rm GC} = {\rm \Pi}\hat{e}_R + {\rm \Theta}\hat{e}_{\phi}+ {\rm Z}\hat{e}_{z}\\
\end{aligned}
\end{equation}
where $(l, b)$ are the Galactic longitude and latitude, $R_\sun = 8.34$~kpc \citep{reid14} is the adopted distance to the GC, $\Pi = \Vr \equiv \dot{R}$, $\Theta = V_{\phi} \equiv R \dot{\phi}$, and $Z = \Vz \equiv \dot{z}$. 

The errors in the velocity components are estimated by Monte Carlo sampling assuming a multivariate Gaussian distribution. For each star, 1000 sets of variables used in the velocity calculation are generated from a six-dimension multivariate Gaussian distribution, $\mathcal{N}(\vec{\mu}, \vec{\Sigma}_{\rm \delta})$, where $\vec{\mu} = [{\rm ra}, {\rm dec}, \mu_{\alpha}, \mu_{\delta}, D, v_{\rm rad}]$ are observed values, and $\vec{\Sigma}_{\rm \delta}$ is the covariance matrix determined from the observational errors and the correlation coefficients provided by the Gaia DR2 catalog. 

The velocity components are listed in Table~\ref{tab:3d_vel}.  Along with these components, we list the heliocentric RV, which was used in Z17 and CD14 as the criterion for LA membership. Specifically, stars with $RV  > 100$~\kms were considered members; in this Table we highlight these values in bold. A total of seven stars fit this criterion. It is enlightening to inspect the cylindrical velocity components of these seven stars: All of them have large $V_{\phi}$,  indicating a strong rotation component. This is unexpected for potential LA members, since the Clouds have a nearly-polar orbit \citep{kall13}. It is the proper motions of these stars that now challenge the notion that these are LA members.

\begin{figure}
\centering
\includegraphics[scale=0.4]{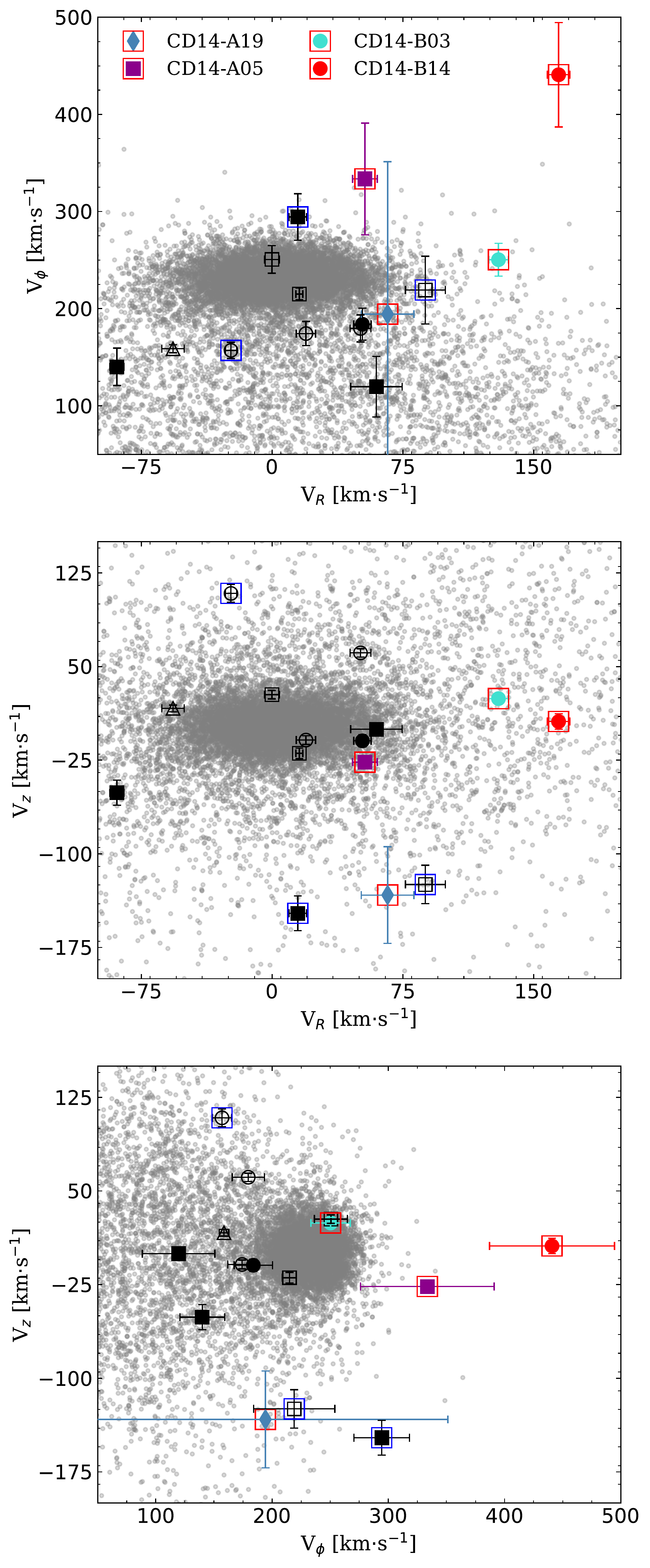}
\caption{The distributions of velocity components. Open and filled symbols represent stars from the present study and from Z17, respectively. Squares, circles and triangle represent stars with “CD14-A**” (region LA I) , “CD14-B**” (region LA II), and “CD14-C**” (region LA III), respectively. The filled diamond is a faint, distant star from the CD14 study which has no high-resolution spectroscopic measurement. This star, CD14-A19, has large errors, due to the its large distance error (50\%). Four G1 stars with large pericenters (i.e., large orbital energies) are marked by surrounding red squares. The G2 stars with large velocity components but moderate orbital energies are marked by surrounding blue squares. Background gray dots show Galactic disc O/B stars \citep{liu18} from the DR5 of the LAMOST survey, with velocity errors $< 50$~\kms. This latter sample is representative of the kinematics of the thin and thick disks.}
\label{fig:vels_3d}
\end{figure}

We plot the velocity components in Figure~\ref{fig:vels_3d}. Together with our target stars, we plot a set of Galactic disk O/B stars. These stars are from the DR5 catalog of the Large Sky Area Multi-Object Fiber Spectroscopic Telescope (LAMOST, also called the Guo Shou Jing Telescope) survey \citep{cui12}. Only stars with velocity errors $<50$~\kms from this catalog \citep{liu18} are shown. We use this sample of O/B stars as representative for the thin/thick disk kinematics. From Figure~\ref{fig:vels_3d} and Table~\ref{tab:3d_vel} it can be seen that eight stars have velocity components well within the range of thin/thick disk stars: low $\Vr$ and $\Vz$ with $V_{\phi} \sim 120 - 250 $ \kms. These stars are CD14-A11, A13, A15, A18, B02, B04, B18, and C06. Three of these stars --- specifically CD14-A15, B02 and B04 --- were previously assumed to be LA members according to their RVs. The remaining seven stars from our sample have peculiar velocities with respect to the canonical thin/thick disk, i.e., with large $\Vr$ or $\Vz$ or $V_{\phi}$. These high-velocity stars are highlighted in Figure~\ref{fig:vels_3d} with blue and red symbol contours; the red contour is for the most energetic stars among these high-velocity stars (see next subsection). We will next focus on the orbits of our fifteen stars to further investigate their origin.

\subsection{Orbit Integration and Parameters}
\label{subsec:orbits}
\begin{figure}
\centering
\includegraphics[scale=0.5]{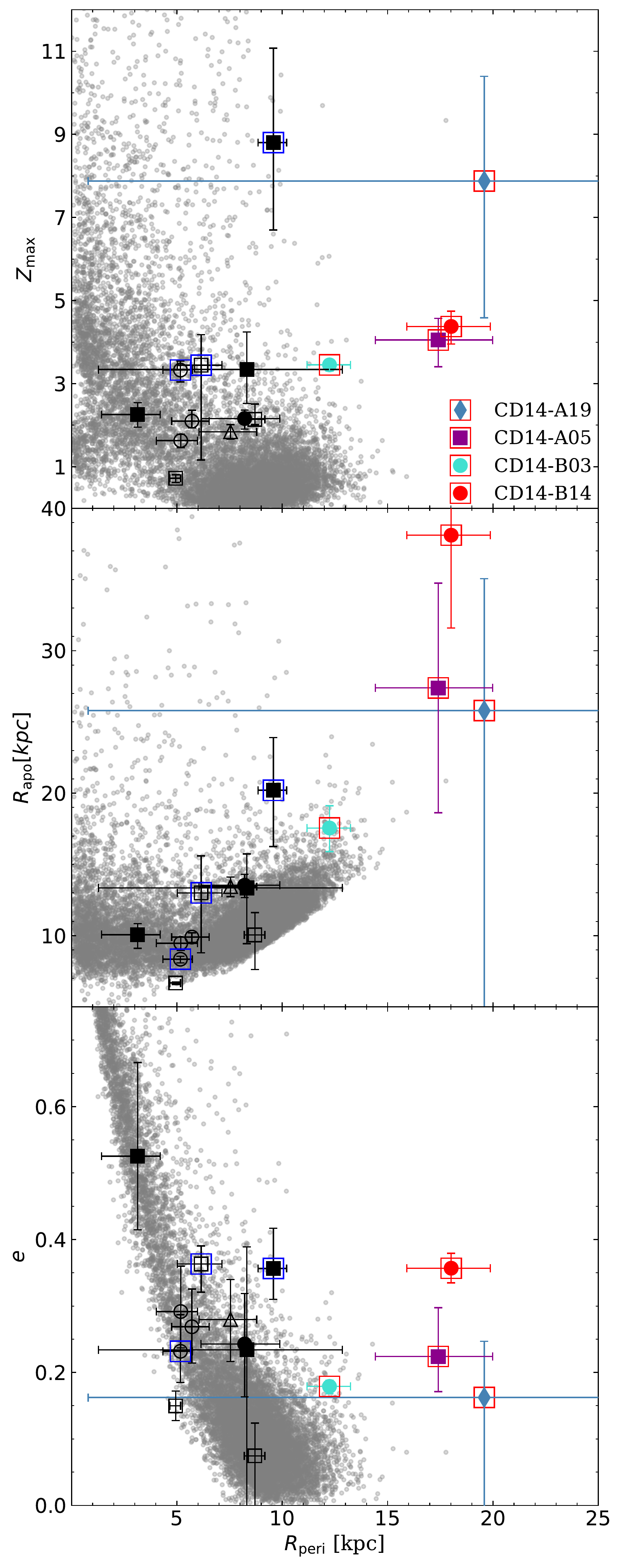}
\caption{Similar to Figure~\ref{fig:vels_3d}, but for orbit parameters.}
\label{fig:orbit_par}
\end{figure}

\begin{figure}
\centering
\includegraphics[width=0.75\textwidth]{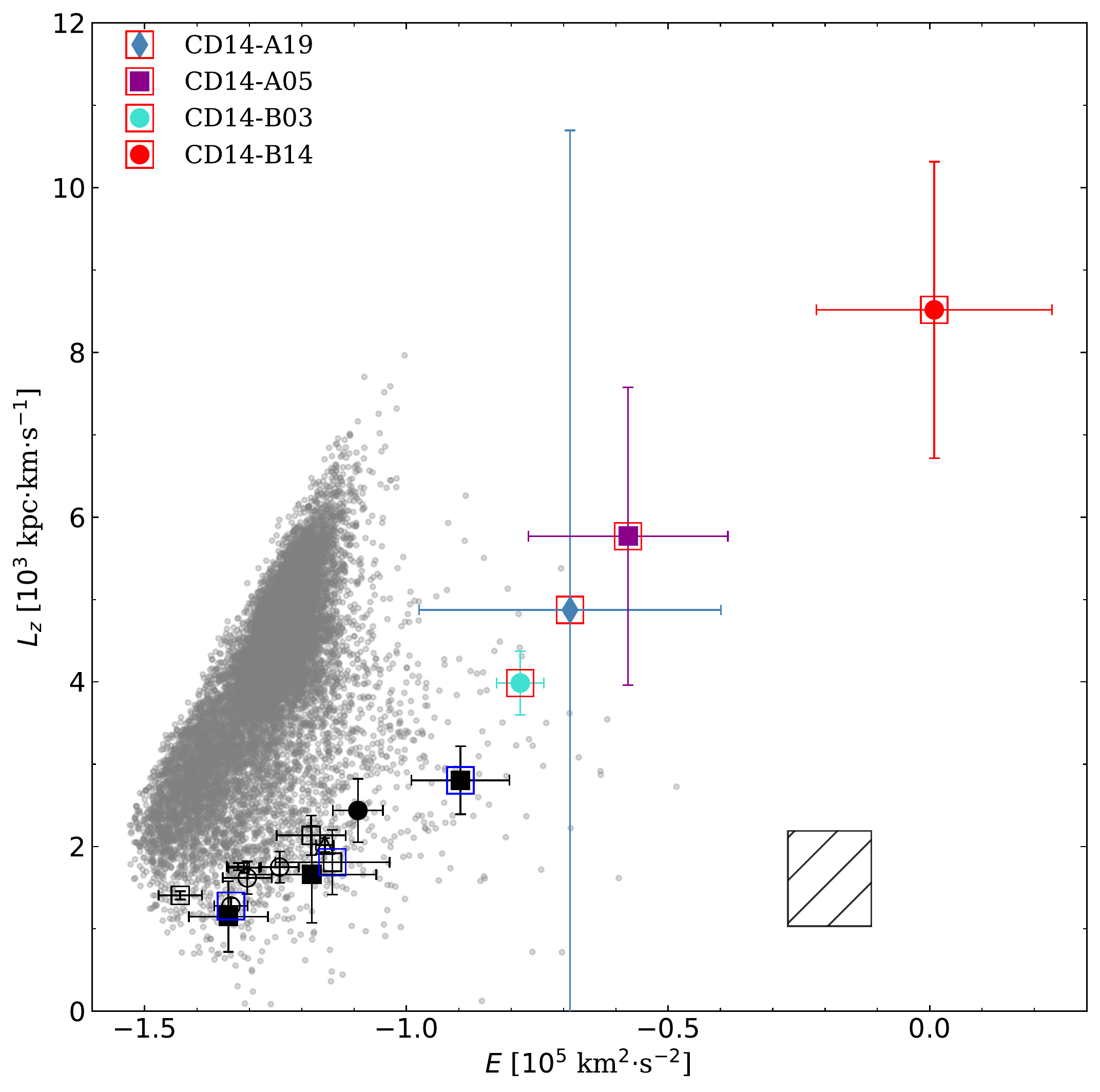}
\caption{The location of the analyzed stars in the energy-angular momentum plane. Symbols have the same meaning as in Figure~\ref{fig:vels_3d}. The hatched area and its size indicate the location of the LMC and the error region, respectively.}
\label{fig:E_Lz}
\end{figure}

Orbit integrations were done via {\tt galpy}\footnote{\url{http://github.com/jobovy/galpy}}, a {\tt python} package for dynamical calculations \citep{bovy15}. In the calculations, the bulge, disk and dark-halo contributions are modeled by a power law with exponential cutoff, a Miyamoto-Nagai disk \citep{miya75}, and a Navarro-Frenk-White potential \citep{nfw97}, respectively. The uncertainties of orbit integrations are also estimated through Monte Carlo sampling. The results for the orbit parameters are listed in Table~\ref{tab:orbit_para} and plotted in Figure~\ref{fig:orbit_par}. Total specific orbital energy $E$ and $L_{z}$ orbital angular momentum are shown in Figure~\ref{fig:E_Lz}. $E$ and $L_z$ of the LMC under the same potential are also shown in this plot. The parameters used in the calculations of $E$ and $L_z$ of the LMC are: $D = 49.97\pm0.19$~kpc \citep{piet13}, $RV = 262.2\pm3.4$~\kms \citep{van02}, $\mu_{\alpha} \cos{\delta} = 1.910\pm0.020$~mas~yr$^{-1}$, and $\mu_{\delta} = 0.229\pm0.047$~mas~yr$^{-1}$ \citep{kall13}.

The orbital parameters of our target stars indicate that a large number of them fit well within the thin/thick disk populations in agreement with the velocity results. The seven high-velocity stars mentioned in the previous subsection are CD14-A05, A07, A12, A19, B03, B14 and B19. Four of these stars are also the most energetic ones in our sample with large pericentric radii, ($R_{\rm peri} > 12$ kpc), and high orbital energies. These are CD14-A05, A19, B03 and B14. We caution that CD14-A19 has large errors due to its large distance error; nevertheless, its orbital energy seems large even within the quoted error (see Figure~\ref{fig:E_Lz}). The remaining three high-velocity stars have pericentric radii less than $\sim 10$ kpc. From Figure~\ref{fig:E_Lz}, they appear to have less rotation ($L_{z}$) at a given energy when compared to thin disk stars (the dense gray plume of objects in Figure~\ref{fig:E_Lz}). In this regard, they fit with the low-rotation tail of the thick disk. Overall, none of our stars have orbital parameters compatible with an association to the LMC (see Figure~\ref{fig:E_Lz}).

%____________________________________________________________________________________
%Section 6:  discussion
%____________________________________________________________________________________
\section{Origin of the Stars}
\label{sec:disc}
\subsection{Orbits}
\label{subsec:kine_orbit_res}
Although in the previous subsections we have touched upon the origin of these stars as indicated from velocities and orbital elements, here we further explore the orbits of the stars in the context of their lifetime. We separate our fifteen stars in the three groups already hinted at in the previous subsections. 
The groups are:
1) Group 1 (G1): four stars with large velocity components and high orbital energies: CD14-A05, A19, B03, B14;
2) Group 2 (G2): three stars with large velocity components, but moderate orbital energies: CD14-A07, A12, B19; and
3) Group 3 (G3): the remaining eight target stars, with no kinematical peculiarities with respect to the Galactic thin/thick disk population. 

\begin{figure}
\gridline{\fig{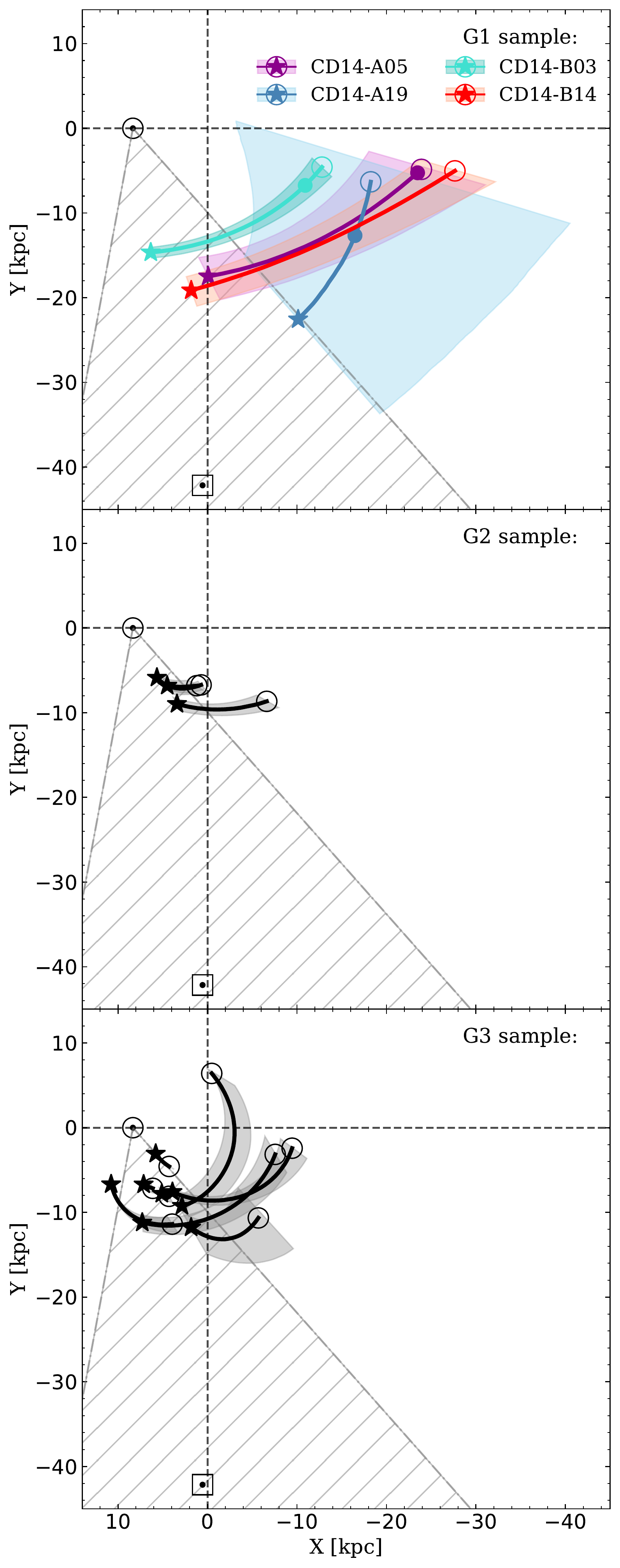}{0.487\textwidth}{(a)}
          \fig{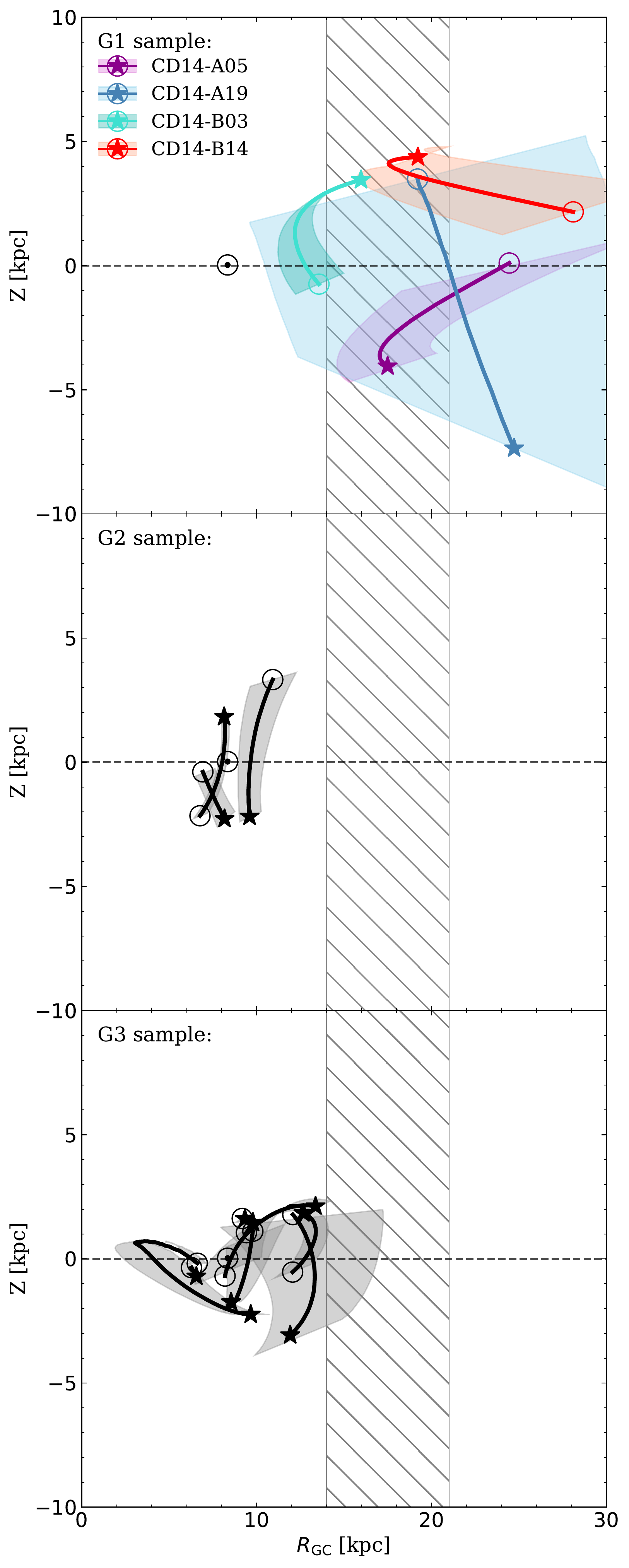}{0.497\textwidth}{(b)}
          }
\caption{Orbits of the target stars in the Galactic X-Y plane (a) and in the $\rgc$-Z plane (b). Filled star symbols and open circle symbols represent current positions and positions of the birthplace, respectively. Shades depict the error range of the orbits. 
Sun's location is represented with a circle-dot symbol, while LMC's location is represented with a square-dot symbol. In the upper panel of a), the filled circles along the trajectories mark the crossing time of the Galactic plane. The hatched areas in panels (a) indicate the observed sky span of the LA \citep{nide10} projected onto the X-Y plane. In panels (b) the hatched areas show the $\rgc$ range of the LA estimated in \citet{mccl09}.}
\label{fig:orbit_integ_space}
\end{figure}

\begin{figure}
\gridline{\fig{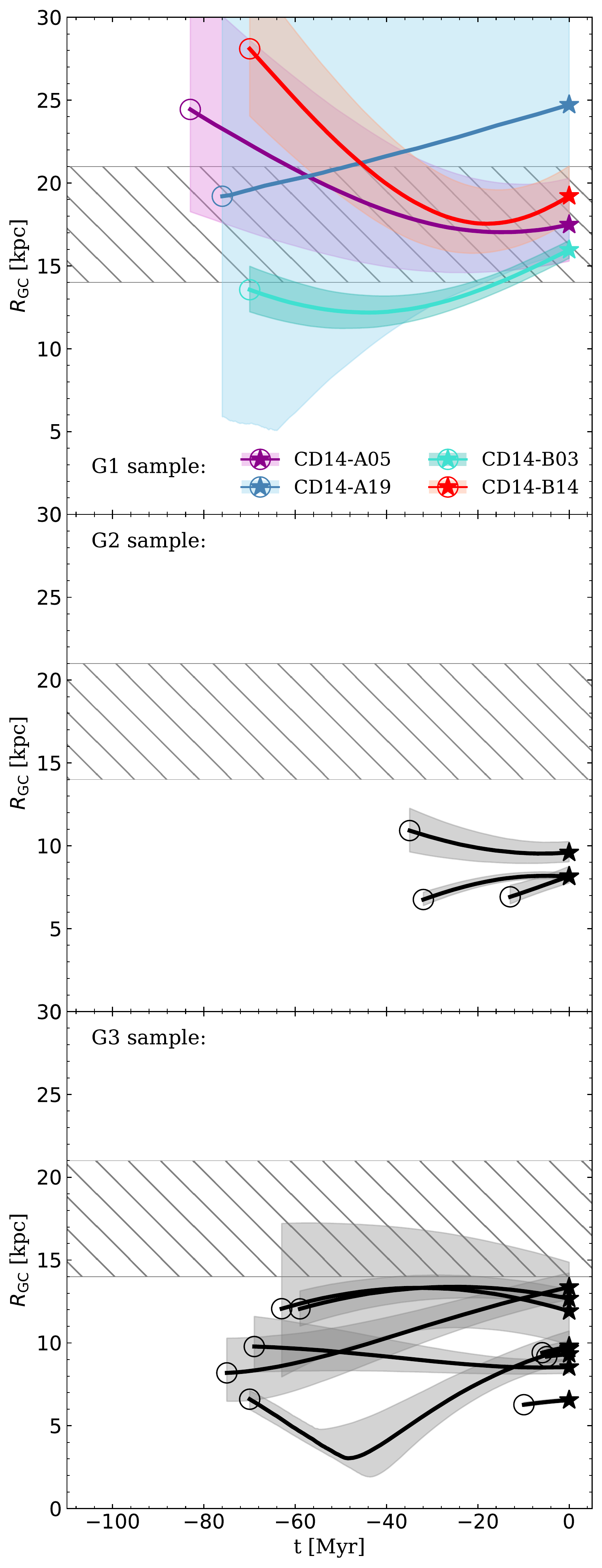}{0.49\textwidth}{(a)}
          \fig{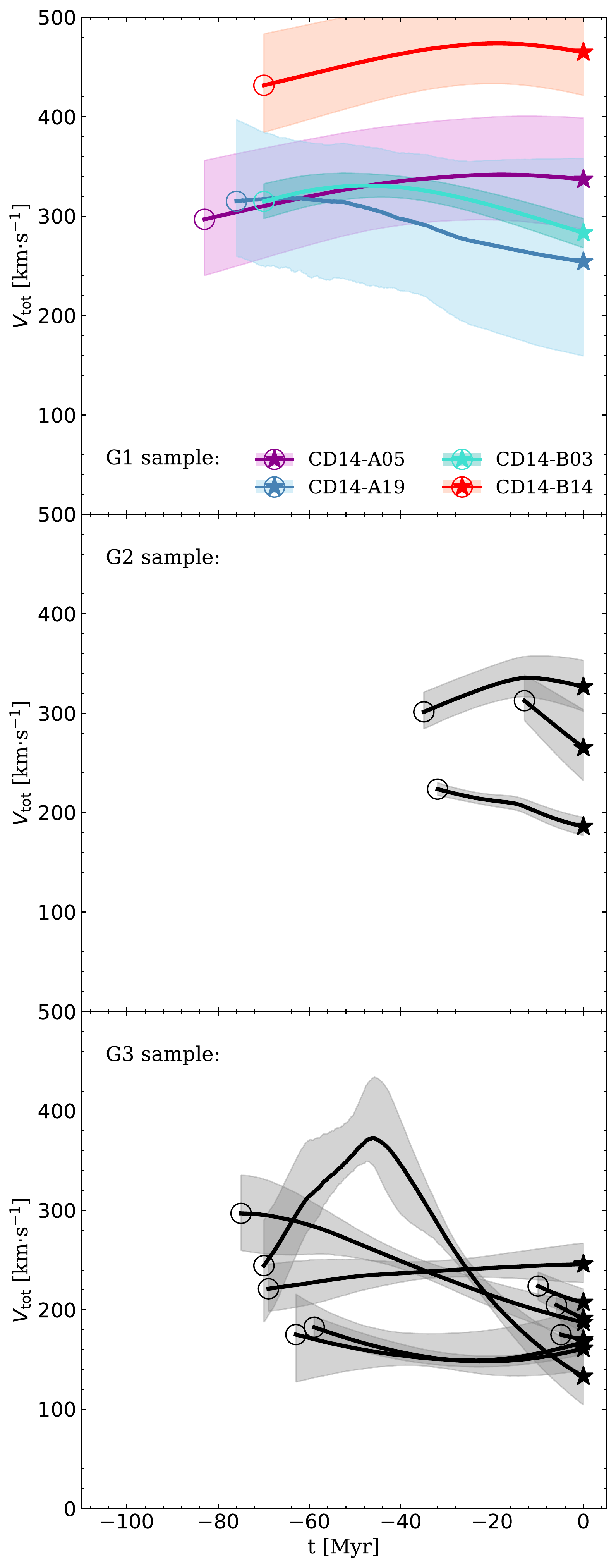}{0.50\textwidth}{(b)}
          }
\caption{Similar to Figure~\ref{fig:orbit_integ_space}, but for variations of $\rgc$ and $\Vtot$ with integration time. The hatched area in plot (a) indicate the $\rgc$ range of the LA estimated in \citet{mccl09}.}
\label{fig:orbit_integ_time}
\end{figure}

The ages are adopted as the integration time. We show the orbits in Figure~\ref{fig:orbit_integ_space}: left panels show the X-Y Galactic plane, and right panels show the $R_{GC} -Z$. The top row shows stars in G1, with each star labeled; the middle row shows stars in G2, and the third row shows stars in G3. In Figure~\ref{fig:orbit_integ_time} we show the $R_{GC}$ (left panels) and the total velocity $V_{tot}$ (right panels) as a function of time. G3 orbits (bottom rows) indicate in-plane, more-or-less circular orbits with $Z$ excursions up to 2.5 kpc. These are typical thin/thick disk like orbits. Their birth place  is not far from the Galactic plane, and within 5 to 12 kpc from the GC. G2 orbits are qualitatively very similar to G3 orbits; total velocity appears to be somewhat larger. Therefore, G2 stars indicate a disk origin as well. The most probable origin of G2 stars and perhaps some G3 stars is in the thin disk, and then scattered into more thick-disk like orbits via dynamical interactions of binary stars in young associations, i.e., disk runaway type \cite[e.g.,][]{brown15, irrg18}. 

\begin{figure}
\centering
\includegraphics[width=0.75\textwidth]{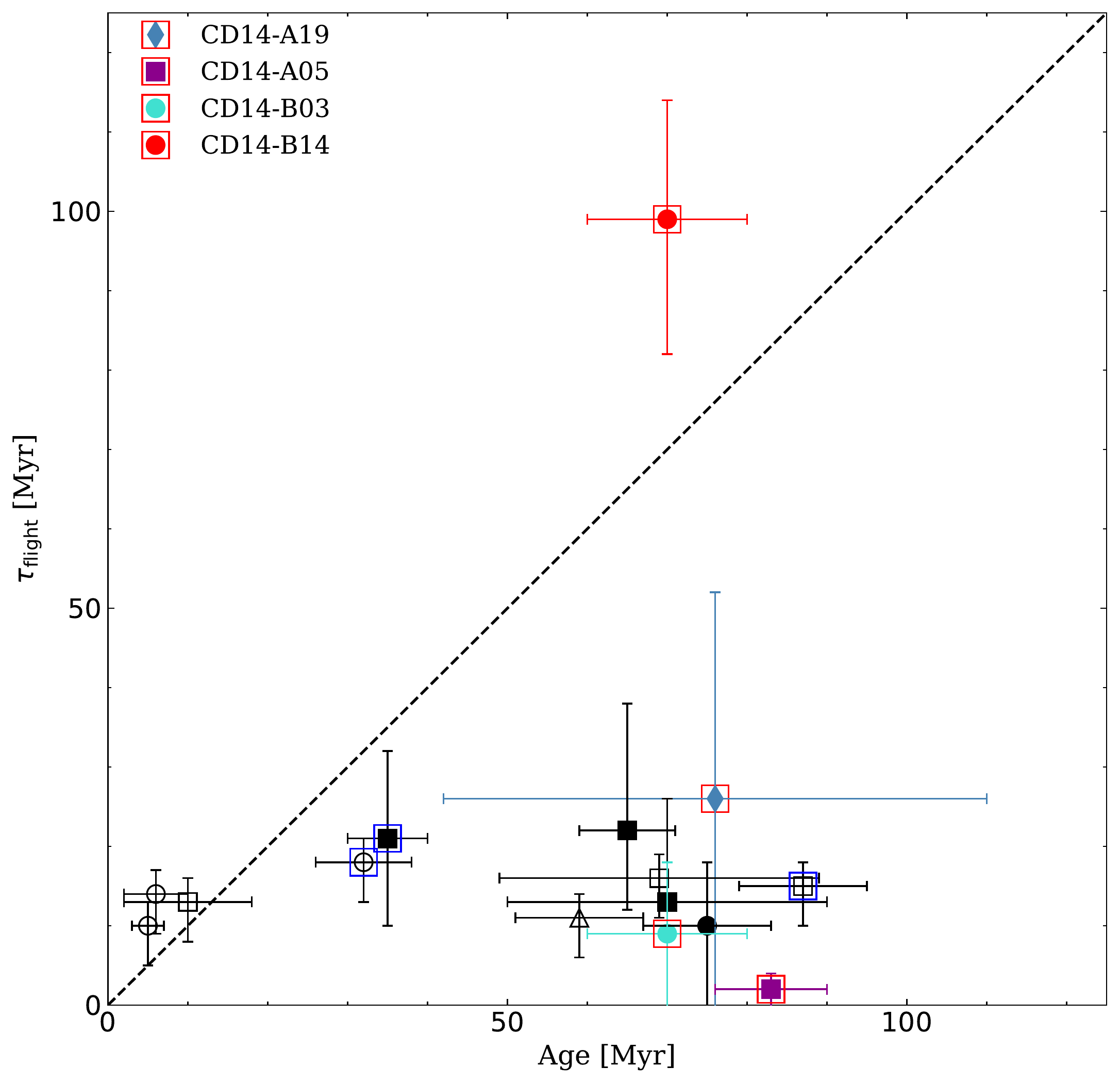}
\caption{The time of flight to reach the Galactic plane versus age for each star. The symbols have the same meanings as Figure~\ref{fig:vels_3d}.}
\label{fig:t_flight}
\end{figure}

We also show times of flight to reach the Galactic plane ($t_{\rm flight}$, listed in Table~\ref{tab:orbit_para}) versus the ages of the stars in Figure~\ref{fig:t_flight}. With the exception of one star, CD14-B14, $t_{\rm flight}$ of G1, G2 and G3 stars are shorter than or comparable to their ages within errors. This also implies a disk origin \citep{irrg18}. Two stars in G3 have typical thin-disk velocities (and thus origin), with no need to invoke any dynamical interactions; these are CD14-A13, and CD14-A18.

G1 stars appear to have originated from near the Galactic plane (except CD14-A19 which has large error bars). They were also born at $\rgc \sim 20$ kpc. The fact that they were born near the galactic plane, and have substantial rotation implies a disk origin as well. Thus their origin appears unrelated to the gaseous material from the LA which should follow the Clouds' orbit about the Galaxy, or a nearly-polar orbit. 
Another possible origin, still related to the Magellanic system, was proposed by
\citet{boub17}. They  proposed a runaway origin from the LMC for some hyper velocity stars, and computed a model to determine the spatial distribution and kinematics of such stars. Our G1 stars do not fit this scenario either. This is because of the G1 stars' large $V_{\phi}$, an aspect also reflected in their large $\mu_l$ proper motions (see our Table~\ref{tab:pms}, values of the order of -4 mas/yr) compared to the predictions ($\sim -2$ mas/yr) in \citet[][their Figure~3]{boub17}.

If G1 stars also have a disk runaway origin, then their birth place is far out in the disk.  Specifically, the plane crossing is between 15 and $\sim 30$ kpc
(see Figure~\ref{fig:orbit_integ_space}). Is this unusual? Among the 15 B-type, hyper velocity stars studied by \citet{irrg18} in light of Gaia DR2 proper motions, only two have an unambiguous disk runaway origin at radii $15 - 20$ kpc, namely HVS7 and B485. Only one star in their sample, namely HVS3, which is also 
the youngest at $\sim 18$ Myr, has an LMC runaway origin. The remaining stars in their sample are either runaways from near the Solar circle, or have too large proper-motion uncertainties for a definite conclusion. Total velocities of our G1 stars are well within the range of disk runaway stars, albeit at the lower end: see our Figure~\ref{fig:orbit_integ_time} and Table~1 in \citet{irrg18} \citep[see also][]{brown18}. Therefore, it does appear that our G1 stars have been born rather far away in the disk, but there is some overlap with the findings of \citet{irrg18}.

Besides their birthplace in the outer disk, another intriguing aspect of our G1 stars is that they all originate from a similar Galactic longitude $l \sim 350\deg$.

This may be a selection effect due to our choice of potential LA members in a given area on the sky and with $RV > 100$~\kms. 

To conclude, the kinematical evidence points to a disk runaway origin for most of the stars in our sample; the most energetic ones were born in the outskirts of the Galactic disk at 12 -- 28 kpc from the GC. Only one star, CD14-B14, has an age shorter than the time to reach the plane, indicating a birthplace some 2.5 kpc above the disk. It is hard to imagine Galactic OB associations exist at $\sim 2.5$ kpc above the plane, and at a distance of $\sim 28$ kpc from the GC. Could this star have an extragalactic origin? If so, from what system, since an association with the Magellanic system seems unlikely. 

\subsection{Abundances}
\label{subsec:kine_abun}
We then investigate abundances as functions of angular momentum and energy for our target stars. As in our previous study,  we choose three clusters to chemically represent the Galactic disk, the LMC and the SMC. These clusters are NGC 4755, NGC 2004, and NGC 330 respectively, with abundances from \citet{trun07}; here we use the average abundance of Mg for B-type stars in these clusters. 

\begin{figure}
\centering
\includegraphics[scale=0.75]{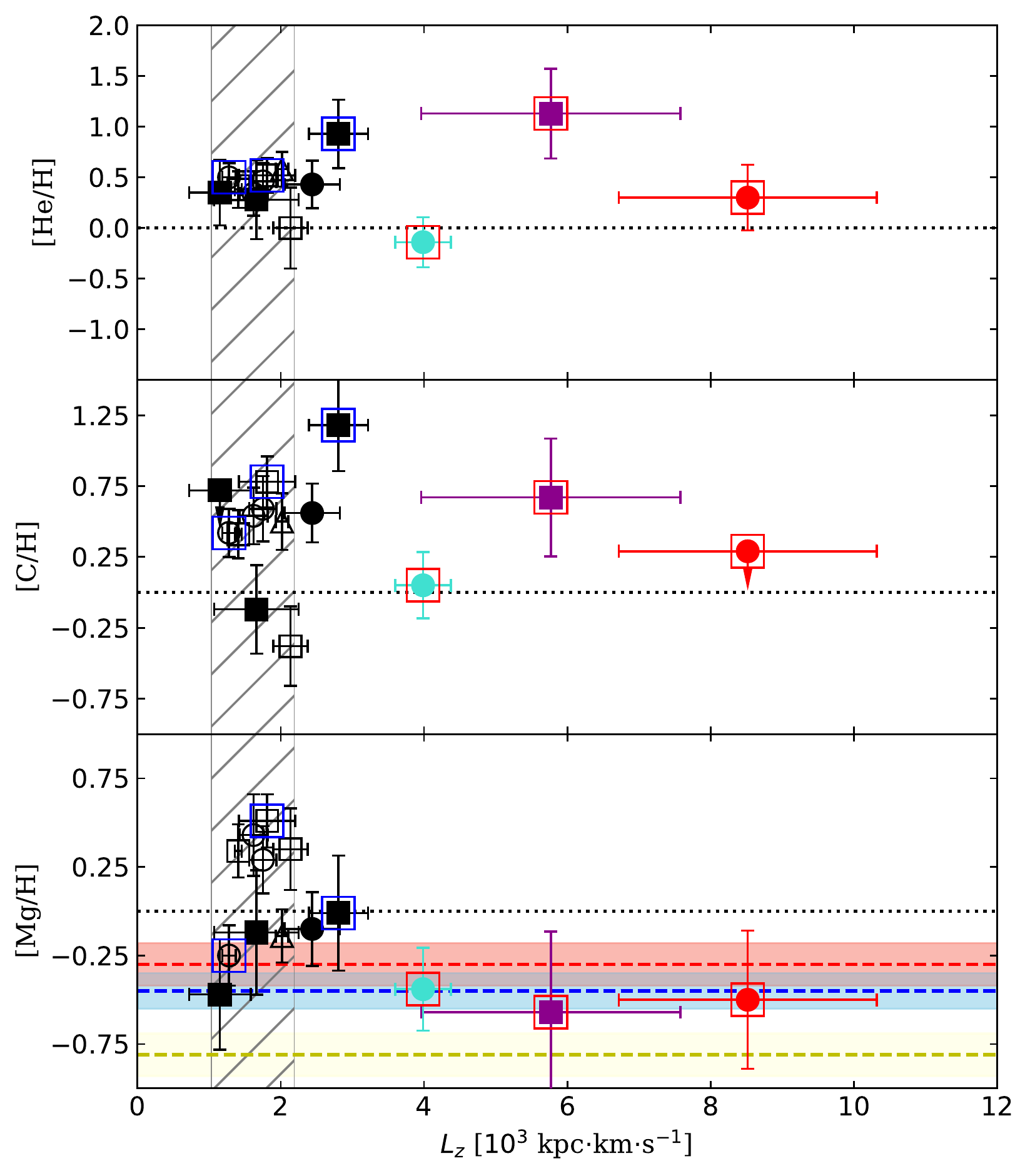}
\caption{Abundance results for He, C, and Mg as a function of angular momentum. Symbols have the same meanings as the ones in Figure~\ref{fig:vels_3d}. The horizontal dashed lines indicate the mean solar abundance of a given element. 
In the bottom panel, red, blue, and yellow dashed lines and their corresponding color-coded shaded areas represent the average [Mg/H] and $1-\sigma$ region of B stars in NGC 4755 (MW), NGC 2004 (LMC), and  NGC 330 (SMC), respectively. The vertical hatched area represents the $L_z$ and its 1-$\sigma$ error range of the LMC.}
\label{fig:Lz_abun}
\end{figure}

\begin{figure}
\centering
\includegraphics[scale=0.75]{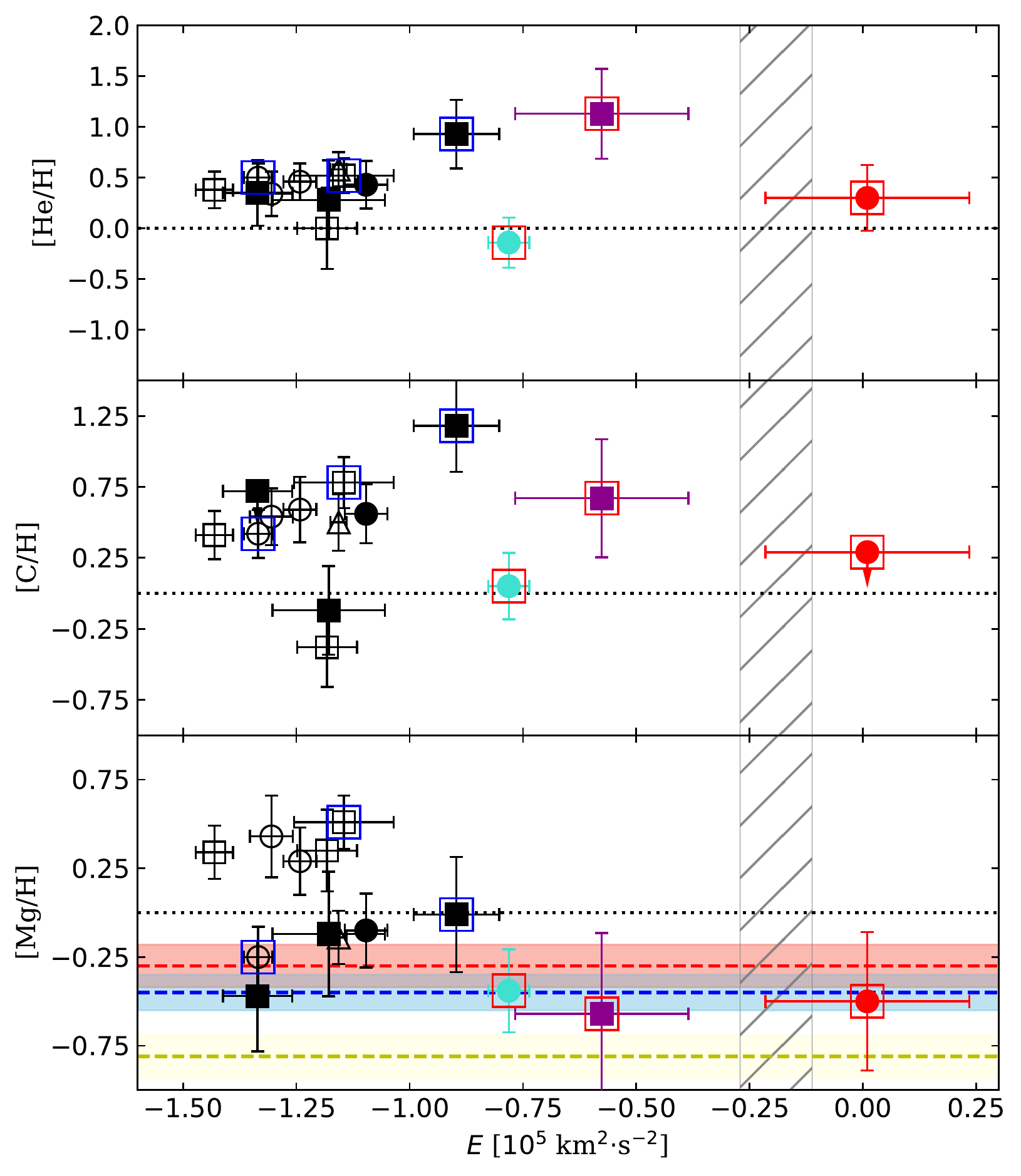}
\caption{Similar to Figure~\ref{fig:Lz_abun}, but for abundance results for He, C, and Mg as a function of total energy.}
\label{fig:E_abun}
\end{figure}

We inspect He, C and Mg, since these are best determined for the entire set of stars. Figures~\ref{fig:Lz_abun} and \ref{fig:E_abun} show the abundance ratios for He, C, and, Mg as functions of angular momentum and energy, respectively. G1 stars are highlighted by surrounding red squares. There are only three stars, since CD14-A19 does not have high-resolution spectroscopic observations. 

In the bottom panels of Figures~\ref{fig:Lz_abun} and \ref{fig:E_abun} we also indicate the average [Mg/H] abundance for our three representative samples; the Galactic disk (red dashed line), the LMC (blue dashed line) and the SMC (yellow dashed line). The color-coded shaded areas correspond to $1-\sigma$ ranges around the averages. 

At high energy and angular momentum Mg is low. Our three G1 stars have an average of [Mg/H] = $-0.50\pm0.07$, which is compatible with LMC's Mg abundance. The remaining stars have an average [Mg/H] = $0.12\pm0.36$. They have a large scatter owed to the fact that they split into two groups at [Mg/H] $\sim 0.4$ and [Mg/H] $\sim -0.2$. Therefore, on average, our most energetic stars are formed from the least Mg-enriched material, which happens to be compatible with LMC's abundance. However, this low Mg abundance can simply reflect the abundance gradient in the disk \citep[see e.g., ][and references therein]{mce17}.

C abundance shows a large scatter over the entire range of energy and angular momentum. He abundance on the other hand, is rather tight ($\sigma_{\rm [He/H]} = 0.15$ ) with super solar values for stars with low energy and angular momentum, and it becomes very scattered ($\sigma_{\rm [He/H]} = 0.72$) at high energy and angular momentum. This scatter in He, and possibly C, may reflect different mechanisms that produce runaway stars.
The two main mechanisms that produce disk runaway stars are: the binary supernova process when the runaway is the secondary star in a binary system where the most massive star undergoes a supernova explosion, and the dynamical ejection process when multibody interactions take place in a cluster/association. As discussed in \citet{mce17} \citep[see also][]{przy08a}, both these processes may affect the surface chemical composition of the runaway star.  However, \citet{mce17} conclude that abundance patterns of C, N, Si, and Mg do not differ in B-type runaway stars from their counterparts in open clusters, instead they attribute the scatter of these elements to the different birthplaces of the runaway stars. Increased He abundance can also be due to mixing between surface and interior in these fast-rotating B-type stars. To summarize, the scatter in He (and possibly C) seen in the most energetic stars can be interpreted at best as a confirmation of their runaway status.

%____________________________________________________________________________________
%Section 7: Summary
%____________________________________________________________________________________
\section{Summary}
\label{sec:sum}
We have measured chemical abundances and RVs for nine B-type stars in the region of the Magellanic LA, based on high resolution spectra taken with the MIKE instrument on the 6.5m Clay telescope. Among the nine stars, two were found to be HB stars based on the stellar parameters, and were excluded from the analysis. We combined the remaining seven young stars with the samples in Z17 and CD14 for a total of 15 stars. For these 15  stars, 3D velocities (using Gaia DR2 proper motions) and orbits were determined. We found that none of the RV candidates to LA membership have orbits compatible with a Magellanic origin. This is owed to their large proper motions along Galactic longitude, as measured by Gaia DR2. To this end, the fact that these stars are {\bf not} members of the LA indicates that the abundance of the LA is mainly SMC-like as consistently determined from the gas studies \citep{lu98, fox18, rich18}.

The majority of the stars in our sample originate as: 1) typical thin disk stars (and chosen as such in our observing program for comparison purposes), and 2) runaway disk with approximately thick-disk like kinematics. The ages of these stars (with one exception) are comparable to or larger than their time since disk crossing, thus confirming their runaway status. However, the most energetic star in our sample, CD14-B14 has however an age (70 Myr) shorter than the time since disk crossing, indicating a birthplace some 2.5 kpc above the Galactic plane and at $\sim 28$ kpc from the GC. With an orbital energy comparable to that of the LMC, but an orbital angular momentum much larger, it is unclear what the origin of this star is. The next three most energetic stars have birthplaces ranging from 12 to 25 kpc from the GC. These four most energetic stars also appear to have been born at a similar Galactic longitude of $\sim 350\deg$; this may be an observational selection effect.

The three most energetic stars that have abundance measurements have low Mg abundance, with an average comparable to the Mg abundance in the LMC (as represented by cluster NGC 2004). This may also simply imply that they formed from less
enriched material in the outskirts of the Galactic disk. The He abundance is tight and super-solar for low orbital-energy stars, while for high orbital-energy stars it is very scattered. This large He abundance scatter may reflect the different mechanisms that produce runaway stars.

%____________________________________________________________________________________
%Acknowledgements
%____________________________________________________________________________________
\acknowledgments
We thank Dr. Z.-X. Lei and Dr. C. Liu for discussions of the stellar evolution, {\bf and the anonymous referee for helpful comments.} This project was developed in part at the 2018 Gaia-LAMOST Sprint workshop, supported by the National Science Foundation of China (NSFC) grants 11333003 and 11390372. L.Z. acknowledges supports from NSFC grants 11773033 and 11673030.  D.I.C. acknowledges partial support from NASA grant 80NSSC18K0422. C.M.B. acknowledges support from project FONDECYT regular \# 1150060. R.A.M. acknowledges support from project FONDECYT regular \# 1170854 and from the CONICYT project BASAL AFB-170002. V.K. acknowledges support from RSF grant N 18-12-00213.

Guoshoujing Telescope (the Large Sky Area Multi-Object Fiber Spectroscopic Telescope, LAMOST) is a National Major Scientific Project built by the Chinese Academy of Sciences. Funding for the project has been provided by the National Development and Reform Commission. LAMOST is operated and managed by the National Astronomical Observatories, Chinese Academy of Sciences.

This work has made use of data from the European Space Agency (ESA) mission Gaia (\url{https://www.cosmos.esa.int/gaia}), processed by the Gaia Data Processing and Analysis Consortium (DPAC, \url{https://www.cosmos.esa.int/web/gaia/dpac/consortium}). Funding for the DPAC has been provided by national institutions, in particular the institutions participating in the {\it Gaia} Multilateral Agreement.

%____________________________________________________________________________________
%References
%____________________________________________________________________________________
\bibliographystyle{aasjournal}
\bibliography{ref}

\clearpage

%____________________________________________________________________________________
%Figures
%____________________________________________________________________________________

\clearpage

\clearpage

\clearpage

\clearpage

\clearpage

\clearpage

\clearpage

\clearpage

\clearpage

\clearpage

\clearpage

\clearpage

\clearpage

\clearpage

\clearpage

\clearpage
%____________________________________________________________________________________
%Tables
%____________________________________________________________________________________
\begin{deluxetable}{lcccccccc}
\centering
%\tabletypesize{\scriptsize}
%\tablecolumns{9}
%\tablewidth{0pc}
\tablecaption{An observation log from Magellan/MIKE\tablenotemark{a}}
\tablehead{\colhead{ID} & \colhead{SPM ID} &\colhead{R.A.} & \colhead{Dec.} & \colhead{V} & \colhead{Start} & \colhead{Instrument} & \colhead{Exp.}  & \colhead{$S/N$\tablenotemark{b}} \\
                  \colhead{}     & \colhead{}     & \colhead{[J2000]} & \colhead{[J2000]} &   \colhead{}  & \colhead{[UT]} & \colhead{} & \colhead{[s]} & \colhead{} }
\startdata
\multirow{2}{*}{CD14-A07} & \multirow{2}{*}{0200040767} & \multirow{2}{*}{11:41:54.77} & \multirow{2}{*}{-78:43:59.99} & \multirow{2}{*}{14.9} &  \multirow{2}{*}{06:27:11} & MIKE - Blue & \multirow{2}{*}{$3600\times1$} & 47 \\
  &  &  & & & & MIKE - Red &  & 79  \\
\multirow{2}{*}{CD14-A13} &  \multirow{2}{*}{0400496347} & \multirow{2}{*}{12:13:19.85} & \multirow{2}{*}{-73:58:45.18} & \multirow{2}{*}{16.0} &  \multirow{2}{*}{04:20:46} &MIKE - Blue & \multirow{2}{*}{$3600\times2$} & 50  \\
  &  &  & & & & MIKE - Red &  & 80  \\     
\multirow{2}{*}{CD14-A18} & \multirow{2}{*}{0690042190} &  \multirow{2}{*}{14:19:59.97}  &  \multirow{2}{*}{-71:58:0.06} &  \multirow{2}{*}{14.0} &  \multirow{2}{*}{07:30:48} & MIKE - Blue & \multirow{2}{*}{$1800\times1$}  & 62 \\
  &  &  & & & & MIKE - Red &  & 92 \\
\multirow{2}{*}{CD14-B04} & \multirow{2}{*}{2890060268} &  \multirow{2}{*}{10:41:17.69}  &  \multirow{2}{*}{-45:07:12.93} &  \multirow{2}{*}{14.3}  &  \multirow{2}{*}{00:17:54} & MIKE - Blue & \multirow{2}{*}{$2100\times1$} & 70 \\ 
 &  & & & & & MIKE - Red &  & 88  \\
\multirow{2}{*}{CD14-B07} & \multirow{2}{*}{2900036493} &  \multirow{2}{*}{10:55:37.37}  &  \multirow{2}{*}{-46:13:04.73} &  \multirow{2}{*}{15.4}  &  \multirow{2}{*}{01:00:31} & MIKE - Blue & \multirow{2}{*}{$3600\times1 $} & 60 \\
  &  &  & & & & MIKE - Red &  & 71 \\
\multirow{2}{*}{CD14-B16}   & \multirow{2}{*}{2320047259} &  \multirow{2}{*}{11:37:27.77}  &  \multirow{2}{*}{-51:26:15.22} &  \multirow{2}{*}{14.5}  &  \multirow{2}{*}{02:05:04} & MIKE - Blue & \multirow{2}{*}{$1200\times2$} & 70  \\
  &  &  & & & & MIKE - Red & & 92 \\
\multirow{2}{*}{CD14-B18}   & \multirow{2}{*}{2320058750} &  \multirow{2}{*}{11:45:33.72}  &  \multirow{2}{*}{-50:49:45.44} &  \multirow{2}{*}{15.0}  &  \multirow{2}{*}{02:56:00} & MIKE - Blue & \multirow{2}{*}{$3600\times1$} & 70 \\
 &  &  & & & & MIKE - Red &  & 90 \\
\multirow{2}{*}{CD14-B19} & \multirow{2}{*}{2930028829} &  \multirow{2}{*}{12:04:39.79}  &  \multirow{2}{*}{-46:23:50.86} &  \multirow{2}{*}{13.5}  &  \multirow{2}{*}{04:00:22} & MIKE - Blue & \multirow{2}{*}{$900\times1$} & 68 \\
  &   & & & & & MIKE - Red &  & 83  \\
  \multirow{2}{*}{CD14-C06} & \multirow{2}{*}{5540048652} &  \multirow{2}{*}{08:59:47.73}  &  \multirow{2}{*}{-24:00:47.19} &  \multirow{2}{*}{14.1}  &  \multirow{2}{*}{23:31:03} & MIKE - Blue & \multirow{2}{*}{$1200\times2 $} & 71 \\
  &   & & & & & MIKE - Red &  & 91  \\
\enddata
\label{tab:obs_log}
\tablenotetext{a} {The observation was taken on 2016 Apr 30. }
\tablenotetext{b}{The $S/N$s were evaluated at 4000 and 5800~{\AA} for blue and red sides, respectively.}
\end{deluxetable}

\clearpage
\begin{deluxetable}{@{\extracolsep{4pt}}lccccccccc@{}}
%\tabletypesize{\scriptsize}
%\tablecolumns{9}
\tablewidth{500pt}
\tabletypesize{\scriptsize}
\tablecaption{The stellar parameters measured by two independent methods for target stars.}
\label{tab:ste_p}
\tablehead{\\
\multirow{2}{*}{ID} & \multicolumn{2}{c}{$T_{\rm eff}$ [K]} & \multicolumn{2}{c}{$\log g$} & \multicolumn{2}{c}{$\log \frac{N_{\rm He}}{N_{\rm H}}$} & \multicolumn{2}{c}{$v\sin i$ [\kms]} \\ 
 \cline{2-3}  \cline{4-5}   \cline{6-7}  \cline{8-9} \\ [-2ex]
\colhead{} & \colhead{I\tablenotemark{a}} & \colhead{II\tablenotemark{b}} & \colhead{I} & \colhead{II} & \colhead{I} & \colhead{II} & \colhead{I} & \colhead{II} }
\startdata
CD14-A07 & $15000\pm900$    & $14642\pm204$ & $4.25\pm0.13$ & $4.13\pm0.06$ & $-0.44\pm0.17$ & $-0.33\pm0.13$ & $125\pm10$ & 130\\
CD14-A13 & $14000\pm1500$   & $12525\pm102$ & $4.23\pm0.20$ & $4.28\pm0.04$ & $-1.00\pm0.40$ & $-0.81\pm0.09$ & $   70\pm8 $   & 70 \\
CD14-A18 & $14800\pm800$    & $14452\pm141$ & $4.40\pm0.10$ & $4.38\pm0.05$ & $-0.62\pm0.18$ & $-0.67\pm0.08$ & $   65\pm10$ & 60 \\
CD14-B04 & $15750\pm900$    & $15975\pm111$ & $4.26\pm0.08$ & $4.31\pm0.04$ & $-0.54\pm0.18$ & $-0.71\pm0.05$ & $120\pm15$  & 110\\
CD14-B07 & $20500\pm950$    & $20211\pm321$ & $4.56\pm0.13$ & $4.63\pm0.05$ & $-1.53\pm0.16$ & $-1.49\pm0.05$ & $   22\pm8 $   & 40 \\
CD14-B16 & $16500\pm750$    & $15874\pm147$ & $4.30\pm0.13$ & $4.49\pm0.03$ & $-2.05\pm0.15$ & $-1.93\pm0.05$ & $   55\pm10$ & 60 \\
CD14-B18 & $15000\pm900$    & $14735\pm177$ & $4.28\pm0.09$ & $4.23\pm0.05$ & $-0.66\pm0.22$ & $-0.87\pm0.08$ & $220\pm15$  & 210\\
CD14-B19 & $16850\pm700$    & $16612\pm189$ & $4.15\pm0.10$ & $4.14\pm0.04$ & $-0.50\pm0.14$ & $-0.77\pm0.06$ & $    15\pm8 $  & 20 \\
CD14-C06 & $15000\pm600$    & $14254\pm108$ & $4.12\pm0.08$ & $4.11\pm0.03$ & $-0.48\pm0.17$ & $-0.50\pm0.06$ & $    55\pm7 $  & 50 \\
\enddata
\tablenotetext{a}{The parameter I are measured by NLTE grids of BSTARS.}
\tablenotetext{b}{The parameter II re measured by an independent grid (LTE for B stars) and analysis code of LINFOR.}
\end{deluxetable}

\clearpage
\rotate
\centering
\begin{deluxetable}{lccccccccc}
%\tablecolumns{10}
%\tablewidth{500pt}
\tabletypesize{\scriptsize} 
\tablecaption{Element abundance results of the targets \tablenotemark{a}} 
\tablehead{  \\[-2ex]
& CD14-A07\tablenotemark{b} & CD14-A13 & CD14-A18\tablenotemark{b}  & CD14-B04\tablenotemark{b}  & CD14-B07 & CD14-B16 & CD14-B18\tablenotemark{b} & CD14-B19 & CD14-C06 } 
\startdata
$\log \epsilon ({\rm He})$ \tablenotemark{c}   & $11.45\pm0.17$ & $10.93\pm0.40$  & $11.31\pm0.18$  & $11.39\pm0.18$ & $10.40\pm0.13$ & $9.88\pm0.15$ & $11.27\pm0.22$ & $11.43\pm0.12$ & $11.51\pm0.17$ \\
$\log \epsilon ({\rm C})$      &    $9.31\pm0.18$ & $   8.15\pm0.28$ & $   8.94\pm0.17$ & $   9.12\pm0.23$ & $   7.84\pm0.12$ & $8.06\pm0.16$ & $  9.07\pm0.20$ & $  8.95\pm0.14$ & $    9.03\pm0.20$ \\
$\log \epsilon ({\rm N})$     & $...$                       & $...$                        &  $...$                       & $...$                        & $   8.06\pm0.13$ & $<7.82$              & $...$                        & $  8.29\pm0.15$ & $...$    \\
$\log \epsilon ({\rm O})$     & $...$                       & $...$                        & $...$                        & $...$                        & $   8.38\pm0.20$ & $<8.83$              & $...$                        & $  9.34\pm0.13$ &  $...$   \\
$\log \epsilon ({\rm Mg})$  &   $8.09\pm0.15$  & $   7.93\pm0.23$ & $7.92\pm0.15$    & $ 7.87\pm0.19$   & $   7.23\pm0.15$ & $7.07\pm0.18$ & $   8.01\pm0.23$ & $  7.25\pm0.11$ & $7.44\pm0.15$ \\
$\log \epsilon ({\rm Si})$     &   $8.01\pm0.18$ & $   7.50\pm0.26$ & $8.00\pm0.16$    & $  7.62\pm0.18$   & $   7.89\pm0.16$ & $7.82\pm0.18$ & $   7.88\pm0.32$ & $ 7.18\pm0.15$ & $7.77\pm0.17$ \\
$\log \epsilon ({\rm P})$      & $...$                       & $...$                        &  $...$                       & $...$                        &   $6.68\pm0.10$ &  $...$     &    $...$             & $...$ & $...$    \\
$\log \epsilon ({\rm S})$      & $<7.98$               & $< 7.33$                & $7.73\pm0.15$    & $<7.43$                  & $   7.40\pm0.13$ & $7.17\pm0.20$ & $...$                        & $  6.98\pm0.15$ & $<7.53$ \\
$\log \epsilon ({\rm Fe})$    &  $...$                       & $...$                        &  $...$                       & $...$                        &   $7.92\pm0.12$ &  $...$     &    $...$             & $7.60\pm0.18$ & $...$    \\     
\hline
${\rm [He/H]}$    & $0.52\pm0.17$  & $ 0.00\pm0.40$  & $0.38\pm0.18$ & $0.46\pm0.18$ & $-0.53\pm0.13$  & $-1.05\pm0.15$ & $0.34\pm0.22$ & $ 0.50\pm0.12$ & $0.58\pm0.17$ \\
${\rm [C/H]}$      & $0.78\pm0.18$  & $-0.38\pm0.28$  & $0.41\pm0.17$ & $0.59\pm0.23$ & $-0.69\pm0.12$  & $-0.47\pm0.16$ & $0.54\pm0.20$ & $ 0.42\pm0.14$ & $0.50\pm0.20$ \\
${\rm [N/H]}$     & $...$                      & $...$                        &  $...$                    & $...$                     & $  0.14\pm0.13$ & $<-0.10$              & $...$                     & $ 0.37\pm0.15$ & $...$                           \\
${\rm [O/H]}$     & $...$                      & $...$                        &  $...$                    & $...$                     & $-0.45\pm0.20$  & $ < 0.00$             & $...$                     & $ 0.51\pm0.13$ & $...$                           \\
${\rm [Mg/H]}$  & $0.51\pm0.15$  & $ 0.35\pm0.23$   & $0.34\pm0.15$  & $0.29\pm0.19$ & $-0.35\pm0.15$  & $-0.51\pm0.18$ & $0.43\pm0.23$ & $-0.25\pm0.11$ & $ -0.14\pm0.15$ \\
${\rm [Si/H]}$     & $0.46\pm0.18$  & $-0.05\pm0.26$  & $0.45\pm0.16$  & $0.07\pm0.18$  & $  0.34\pm0.16$ & $ 0.27\pm0.18$ & $0.33\pm0.32$ & $-0.37\pm0.15$ & $  0.22\pm0.17$ \\
${\rm [P/H]}$     & $...$                       & $...$                        &  $...$                       & $...$                        & $1.23\pm0.10$ &  $...$     &    $...$               & $...$ & $...$    \\
${\rm [S/H]}$      & $<0.65$               & $<0.00$                & $0.40\pm0.15$ &  $<0.10$              & $  0.07\pm0.13$ & $-0.16\pm0.20$ &  $...$                   & $-0.35\pm0.15$ & $<0.2$ \\ 
${\rm [Fe/H]}$    & $...$         & $...$           &  $...$            & $...$             & $0.42\pm0.12$ &  $...$     &    $...$               & $0.05\pm0.18$ & $...$  \\
\enddata
\label{tab:abun_results_a}
\tablenotetext{a}{Solar compositions are taken from Grevesse \& Sauval (1998).}
\tablenotetext{b}{With $Z/Z_{\odot}=2$ model atmosphere.}
\tablenotetext{c}{$\log \epsilon ({\rm X}) = \log \frac{N_{\rm X}}{N_{\rm H}} + 12$.}
\end{deluxetable}

\clearpage
\begin{deluxetable}{@{\extracolsep{4pt}}llccccrr@{}}
%\tabletypesize{\scriptsize}
%\tablecolumns{7}
%\tablewidth{0pc}
\tablecaption{Distance, age, and radial velocities}
\tablehead{\\
\multirow{2}{*}{ID} &  & \multicolumn{2}{c}{$(m-M)_0$\tablenotemark{c} (mag)} & \multicolumn{2}{c}{Age} & \multicolumn{2}{c}{RV [\kms]} \\ [1ex]
 \cline{3-4}  \cline{5-6}   \cline{7-8}  \\ [-2ex]
\colhead{} & \colhead{} & \colhead{MS} & \colhead{HB} & \colhead{MS   (Myr)} & \colhead{HB  (Gyr)} & \colhead{TW\tablenotemark{d}} & \colhead{CD14} }
\startdata
CD14-A07\tablenotemark{a} & & $14.54\pm0.30$ & $...$ & $ 13\pm12$  & $...$ & $ 94\pm5$   & $ 87\pm8$ \\
%CD14-A13                                       &  & $15.20\pm0.47$ & $...$                       & $ 45\pm40$ & $...$    & $-2\pm7$    & $-12\pm6$ \\ 
\multirow{2}{*}{CD14-A13}      & SA~I   & $15.20\pm0.47$ & $...$    & $ 45\pm40$ & $...$    & \multirow{2}{*}{$-2\pm7$}    & \multirow{2}{*}{$-12\pm6$} \\
                                                                    & SA~II & $14.76\pm0.21$ &  $...$    & $ 69\pm20$ & $...$    &       &  \\
CD14-A18\tablenotemark{a} & & $13.05\pm0.28$\tablenotemark{b} & $...$ & $ 10\pm8$\tablenotemark{b}  & $...$ & $-20\pm5$  & $-20\pm6$  \\ 
CD14-B04\tablenotemark{a} & & $14.21\pm0.22$ & $...$ & $ 6\pm4$    & $...$ & $132\pm4$  & $136\pm5$ \\ 
%CD14-B07                               & $15.33 \pm0.26$& $12.62\pm0.50$ & $<0.01$                & $> 3$  & $19\pm4$   & $25\pm7$  \\ 
%CD14-B16                               & $14.34\pm0.29$ & $11.62\pm0.40$ & $ 10\pm13$  & $> 3$ & $9\pm5$      & $ 7 \pm5$  \\ 
CD14-B07                  & & $...$ & $12.62\pm0.50$ & $...$  & $> 3$  & $19\pm4$   & $25\pm7$  \\ 
CD14-B16                  & & $...$ & $11.62\pm0.40$ & $...$  & $> 3$  & $9\pm5$    & $ 7\pm5$  \\ 
CD14-B18\tablenotemark{a} & & $14.67\pm0.23$ & $...$ & $ 5\pm2$  & $...$ & $92\pm9$    & $81\pm3$  \\ 
CD14-B19                                & & $14.13\pm0.23$ & $...$                       & $ 32\pm6$     & $...$   & $96\pm4$     & $111\pm7$  \\ 
CD14-C06                                & & $14.33\pm0.18$ & $...$                       & $ 59\pm8$   & $...$    & $96\pm4$    & $99\pm6$ \\ 
\enddata
\label{tab:rv_dis_age}
\tablenotetext{a}{PARSEC isochrones with metallicity of $Z/Z_{\odot}=2$ are used.}
\tablenotetext{b}{$M_v$ of CD14-A18 is assumed at the same temperature from the ZAMS.}
\tablenotetext{c}{$A_{V}$ is calculated assuming an extinction to reddening ratio $A_V/E(B-V) = 3.1$. $E(B-V)$ is from \citet{schlafly11}.}
\tablenotetext{d}{The Heliocentric radial velocity values calculated in this work.}
\end{deluxetable}

\clearpage
\rotate
\begin{deluxetable}{lrrrrrrrrrr}
\tablewidth{400pt}
\tabletypesize{\scriptsize} 
\tablecaption{Proper motions and parallaxes cross-matched from Gaia DR2}
\tablehead{\\
\colhead{ID} & \colhead{Gaia ID} & \colhead{RV} &\colhead{Distance\tablenotemark{a}} & \colhead{plx} & \colhead{$\mu_{\rm ra}$} & \colhead{$\mu_{\rm dec}$} & \colhead{$l$} & \colhead{$b$} & \colhead{$\mu_l$} & \colhead{$\mu_b$} \\ [-2ex]
\colhead{} & \colhead{} & \colhead{[\kms]} & \colhead{[kpc]} & \colhead{[mas]} & \colhead{[mas/yr]} & \colhead{[mas/yr]}  & \colhead{[deg]} & \colhead{[deg]} & \colhead{[mas/yr]} & \colhead{[mas/yr]} }
\startdata
CD14-A05\tablenotemark{b} & 5226326178138343168 & $133\pm6$ & $19.95\pm 2.96$ & $ 0.025\pm0.033$ & $-4.32\pm0.07$ & $ 1.74\pm0.05$ & 295.75 & -11.78 & $-4.66\pm0.04$ & $-0.06\pm0.07$ \\
CD14-A07                  & 5200092659627780352 & $ 94\pm8$ & $ 8.09\pm 1.20$ & $ 0.072\pm0.036$ & $-8.52\pm0.06$ & $-0.51\pm0.05$ & 299.44 & -16.33 & $-8.04\pm0.05$ & $-2.85\pm0.07$ \\
CD14-A08\tablenotemark{b} & 5224454951082866048 & $ 48\pm8$ & $    ...      $ & $ 0.202\pm0.034$ & $-5.69\pm0.06$ & $-0.07\pm0.06$ & 298.79 & -13.92 & $-5.46\pm0.04$ & $-1.60\pm0.07$ \\
CD14-A11\tablenotemark{b} & 5224985946480348416 & $ 71\pm4$ & $13.80\pm 3.57$ & $ 0.010\pm0.028$ & $-2.91\pm0.05$ & $ 0.05\pm0.04$ & 298.96 & -12.88 & $-2.84\pm0.04$ & $-0.66\pm0.06$ \\
CD14-A12\tablenotemark{b} & 5226533985831636480 & $ 33\pm7$ & $10.47\pm 1.01$ & $ 0.090\pm0.025$ & $-7.21\pm0.05$ & $-1.35\pm0.04$ & 298.81 & -12.09 & $-6.67\pm0.04$ & $-3.03\pm0.05$ \\
CD14-A13                  & 5841494623121376512 & $ -2\pm6$ & $ 8.95\pm 0.86$ & $ 0.003\pm0.040$ & $-6.17\pm0.07$ & $ 1.44\pm0.06$ & 300.26 & -11.30 & $-6.32\pm0.06$ & $ 0.49\pm0.07$ \\
CD14-A15\tablenotemark{b} & 5838307413790168960 & $162\pm6$ & $10.96\pm 1.62$ & $ 0.017\pm0.033$ & $-4.88\pm0.06$ & $ 1.15\pm0.05$ & 300.74 & -11.82 & $-4.99\pm0.05$ & $ 0.52\pm0.05$ \\
CD14-A18                  & 5797716193279770496 & $-20\pm6$ & $ 4.07\pm 0.60$ & $ 0.077\pm0.030$ & $-6.24\pm0.04$ & $-3.93\pm0.04$ & 309.74 & -10.28 & $-7.21\pm0.05$ & $-1.57\pm0.03$ \\
CD14-A19\tablenotemark{c} & 5793052064945812480 & $234\pm6$ & $30.20\pm15.30$ & $ 0.101\pm0.065$ & $-2.18\pm0.19$ & $-1.62\pm0.11$ & 309.35 & -14.09 & $-2.65\pm0.22$ & $-0.61\pm0.03$ \\
CD14-B02\tablenotemark{b} & 5414441420674317184 & $170\pm4$ & $11.48\pm 1.11$ & $ 0.034\pm0.029$ & $-3.65\pm0.04$ & $ 1.47\pm0.04$ & 275.29 &  10.56 & $-3.84\pm0.01$ & $-0.89\pm0.06$ \\
CD14-B03\tablenotemark{b} & 5368186306521653248 & $227\pm5$ & $15.14\pm 0.71$ & $ 0.093\pm0.040$ & $-4.10\pm0.06$ & $ 1.92\pm0.06$ & 277.72 &  13.06 & $-4.50\pm0.02$ & $-0.48\pm0.08$ \\
CD14-B04                  & 5366928533939460480 & $132\pm5$ & $ 6.95\pm 0.67$ & $ 0.128\pm0.038$ & $-4.70\pm0.06$ & $ 3.61\pm0.06$ & 280.11 &  11.91 & $-5.86\pm0.02$ & $ 0.85\pm0.08$ \\
CD14-B07                  & 5363434938821203712 & $ 19\pm4$ & $ 3.34\pm 0.87$ & $ 0.238\pm0.048$ & $-4.44\pm0.08$ & $ 0.20\pm0.07$ & 282.90 &  12.10 & $-4.08\pm0.04$ & $-1.77\pm0.10$ \\
CD14-B14\tablenotemark{b} & 5373677473750021760 & $202\pm5$ & $20.89\pm 2.02$ & $-0.003\pm0.036$ & $-5.46\pm0.05$ & $ 1.42\pm0.05$ & 288.78 &  12.20 & $-5.62\pm0.04$ & $-0.45\pm0.07$ \\
CD14-B16                  & 5369739194904171520 & $  9\pm5$ & $ 2.11\pm 0.43$ & $ 0.072\pm0.035$ & $-7.82\pm0.05$ & $ 2.04\pm0.04$ & 291.36 &   9.76 & $-8.07\pm0.03$ & $-0.29\pm0.05$ \\
CD14-B18                  & 5369992529252768000 & $ 92\pm3$ & $ 8.59\pm 0.83$ & $ 0.032\pm0.038$ & $-5.37\pm0.05$ & $ 0.59\pm0.04$ & 292.43 &  10.69 & $-5.34\pm0.04$ & $-0.81\pm0.05$ \\
CD14-B19                  & 6143258509937783424 & $ 96\pm7$ & $ 6.70\pm 0.65$ & $ 0.081\pm0.049$ & $-4.11\pm0.06$ & $ 3.32\pm0.04$ & 294.58 &  15.72 & $-4.66\pm0.05$ & $ 2.49\pm0.04$ \\
CD14-C06                  & 5654350872120057856 & $ 96\pm6$ & $ 7.34\pm 0.71$ & $-0.008\pm0.045$ & $-0.52\pm0.07$ & $ 0.01\pm0.05$ & 249.94 &  14.26 & $-0.34\pm0.01$ & $-0.40\pm0.08$ \\
\enddata
\label{tab:pms}
\tablenotetext{a} {Spectroscopic distance determined in the present study and Z17.}
\tablenotetext{b} {Sample stars from Z17}
\tablenotetext{c} {Sample star from CD14}
\end{deluxetable}

\clearpage
\begin{deluxetable}{lrrrr}
%\centering
\tablecaption{Velocity components in the Galactic cylindrical coordinate system}
\tablehead{\\
\colhead{ID} & \colhead{$V_R$ [\kms]} & \colhead{$V_{\phi}$ [\kms]} & \colhead{$V_z$ [\kms]} & \colhead{$RV$ [\kms]}}
\startdata
CD14-A05\tablenotemark{a} &   53.4$\pm$7.3  &  332.7$\pm$60.1  &  -26.2$\pm$ 5.5 & {\bf 133$\pm$9} \\
CD14-A07 &   88.4$\pm$11.0 &  218.5$\pm$33.8  & -125.1$\pm$14.9 & 94$\pm$ 5\\
%CD14-A08\tablenotemark{a} &  -19.5$\pm$1.6  &  177.3$\pm$7.9   &  -41.6$\pm$2.4 & 48$\pm$ 8\\
CD14-A11\tablenotemark{a} &  -88.6$\pm$3.9  &  139.9$\pm$18.8  &  -50.9$\pm$10.0 & 71$\pm$10 \\
CD14-A12\tablenotemark{a} &   14.5$\pm$4.7  &  294.2$\pm$23.2  & -147.1$\pm$13.6 & 33$\pm$ 7\\
CD14-A13 &    0.6$\pm$4.4  &  249.8$\pm$13.8  &   27.1$\pm$3.4 & -2$\pm7$ \\
CD14-A15\tablenotemark{a} &   59.1$\pm$14.6 &  118.8$\pm$31.7  &   -0.3$\pm$4.6 & {\bf 162$\pm$8} \\
CD14-A18 &   15.8$\pm$1.9  &  215.1$\pm$6.1   &  -19.5$\pm$4.2 & -20$\pm$5 \\
CD14-A19\tablenotemark{b} &   66.2$\pm$16.0 &  196.5$\pm$161.6 & -133.1$\pm$38.9 & {\bf 234$\pm$6}\\
CD14-B02\tablenotemark{a} &   51.7$\pm$5.1  &  184.7$\pm$17.9  &   -9.5$\pm$5.1 & {\bf 170$\pm$5}  \\
CD14-B03\tablenotemark{a} &  129.4$\pm$5.3  &  250.6$\pm$15.7  &   23.9$\pm$4.4 & { \bf 227$\pm$6} \\
CD14-B04 &   51.1$\pm$5.9  &  179.7$\pm$13.7  &   61.3$\pm$3.4 & {\bf 132$\pm4$} \\
%CD14-B07 &  -36.0$\pm$4.2  &  210.3$\pm$7.0   &  -15.6$\pm$6.6 & 19$\pm$4\\
CD14-B14\tablenotemark{a} &  164.1$\pm$6.8  &  435.4$\pm$51.9  &    6.1$\pm$6.3 & {\bf 202$\pm$7}  \\
%CD14-B16 &    8.7$\pm$4.0  &  224.4$\pm$5.0   &    4.8$\pm$1.1 & 9$\pm$5 \\
CD14-B18 &   19.0$\pm$5.4  &  172.5$\pm$12.3  &   -8.6$\pm$3.4 & 92$\pm$9\\
CD14-B19 &  -23.4$\pm$3.4  &  156.6$\pm$8.3   &  108.4$\pm$7.6 & 96$\pm$4 \\
CD14-C06 &  -56.8$\pm$6.3  &  158.6$\pm$4.0   &   16.4$\pm$2.8  & 96$\pm$4 \\
\enddata
\label{tab:3d_vel}
\tablenotetext{a} {Sample stars from Z17}
\tablenotetext{b} {Sample star from CD14}
\end{deluxetable}

\clearpage
\begin{deluxetable}{lccrrrc}
%\centering
\tablecaption{Orbit parameters}
\tablehead{\\
\colhead{ID} & \colhead{$Z_{\rm max}$ [kpc]} & \colhead{$e$} & \colhead{$R_{\rm apo}$ [kpc]} & \colhead{$R_{\rm peri}$ [kpc]} & \colhead{Age [Myr]} & \colhead{$t_{\rm flight}$ [Myr]\tablenotemark{a}}}
\startdata
CD14-A05\tablenotemark{b} & $4.05^{+0.64}_{-0.52}$ & $0.22^{+0.05}_{-0.07}$ & $27.39^{+8.77}_{-7.35} $ & $17.40^{+2.99}_{-2.58}$ & $83\pm7$ & $2^{+12}_{-2}$\\
CD14-A07 & $3.45^{+2.28}_{-0.73}$ & $0.36^{+0.04}_{-0.03}$ & $13.00^{+4.21}_{-2.60} $ & $ 6.15^{+1.13}_{-0.99}$ & $87\pm8$ & $15\pm14$ \\
%CD14-A08\tablenotemark{a} & $1.21^{+0.02}_{-0.02}$ & $0.28^{+0.03}_{-0.03}$ & $ 7.46^{+0.01}_{-0.01} $ & $ 4.22^{+0.32}_{-0.29}$ & $...$\\
CD14-A11\tablenotemark{b} & $3.34^{+0.82}_{-0.90}$ & $0.23^{+0.25}_{-0.15}$ & $13.35^{+3.92}_{-2.39} $ & $ 8.32^{+7.05}_{-4.53}$ & $65\pm6$ & $22^{+10}_{-16}$\\
CD14-A12\tablenotemark{b} & $8.80^{+2.10}_{-2.27}$ & $0.36^{+0.05}_{-0.06}$ & $20.20^{+3.95}_{-3.71} $ & $ 9.58^{+0.72}_{-0.63}$ & $35\pm5$ & $21\pm11$\\
CD14-A13 & $2.14^{+0.13}_{-0.37}$ & $0.07^{+0.08}_{-0.05}$ & $10.05^{+2.43}_{-1.57} $ & $ 8.70^{+0.50}_{-0.48}$ & $69\pm20$ & $16^{+23}_{-16}$\\
CD14-A15\tablenotemark{b} & $2.26^{+0.30}_{-0.29}$ & $0.53^{+0.11}_{-0.14}$ & $10.07^{+0.95}_{-0.78} $ & $ 3.14^{+1.71}_{-1.08}$ & $70\pm20$ & $13^{+22}_{-13}$\\
CD14-A18 & $0.72^{+0.10}_{-0.08}$ & $0.15^{+0.02}_{-0.02}$ & $ 6.68^{+0.10}_{-0.06} $ & $ 4.94^{+0.28}_{-0.23}$ & $10\pm8$ & $14^{+12}_{-10}$\\
CD14-A19\tablenotemark{c} & $7.88^{+3.29}_{-2.52}$ & $0.16^{+0.41}_{-0.08}$ & $25.80^{+21.57}_{-9.26} $ & $19.59^{+18.80}_{-15.13}$ & $76\pm34$ & $26^{+40}_{-26}$\\
CD14-B02\tablenotemark{b} & $2.16^{+0.25}_{-0.21}$ & $0.24^{+0.08}_{-0.08}$ & $13.54^{+ 0.87}_{-0.77} $ & $ 8.22^{+2.07}_{-1.66}$ & $75\pm8$ & $10^{+13}_{-8}$\\
CD14-B03\tablenotemark{b} & $3.45^{+0.15}_{-0.14}$ & $0.18^{+0.01}_{-0.01}$ & $17.56^{+ 1.66}_{-1.57} $ & $12.24^{+1.06}_{-0.99}$ & $70\pm10$ & $9^{+15}_{-9}$\\
CD14-B04 & $2.10^{+0.16}_{-0.26}$ & $0.27^{+0.05}_{-0.06}$ & $ 9.90^{+ 0.37}_{-0.30} $ & $ 5.71^{+0.96}_{-0.81}$ & $ 6\pm4$ & $14^{+9}_{-6}$\\
%CD14-B07 & $0.80^{+0.24}_{-0.19}$ & $0.19^{+0.02}_{-0.02}$ & $ 8.64^{+ 0.30}_{-0.20} $ & $ 5.94^{+0.32}_{-0.35}$ & $> 3\times10^3$\\
CD14-B14\tablenotemark{b} & $4.38^{+0.42}_{-0.37}$ & $0.36^{+0.02}_{-0.02}$ & $38.11^{+ 6.51}_{-5.72} $ & $18.02^{+2.10}_{-1.86}$ & $70\pm10$ & $99^{+17}_{-15}$ \\
%CD14-B16 & $0.38^{+0.07}_{-0.06}$ & $0.10^{+0.02}_{-0.02}$ & $ 7.84^{+ 0.05}_{-0.04} $ & $ 6.36^{+0.32}_{-0.31}$ & $> 3\times10^3$\\
CD14-B18 & $1.63^{+0.17}_{-0.15}$ & $0.29^{+0.05}_{-0.07}$ & $ 9.47^{+ 0.52}_{-0.43} $ & $ 5.18^{+1.16}_{-0.79}$ & $ 5\pm2$ & $10^{+5}_{-3}$\\
CD14-B19 & $3.33^{+0.28}_{-0.21}$ & $0.23^{+0.05}_{-0.05}$ & $ 8.35^{+ 0.26}_{-0.21} $ & $ 5.17^{+0.84}_{-0.56}$ & $32\pm6$ & $18^{+11}_{-8}$\\
CD14-C06 & $1.84^{+0.17}_{-0.17}$ & $0.28^{+0.06}_{-0.06}$ & $13.45^{+ 0.71}_{-0.67} $ & $ 7.54^{+1.50}_{-1.24}$ & $59\pm8$ & $11^{+14}_{-11}$\\
\enddata
\label{tab:orbit_para}
\tablenotetext{a} {The time of flight to reach the Galactic plane.}
\tablenotetext{b} {Sample stars from Z17}
\tablenotetext{c} {Sample star from CD14}
\end{deluxetable}



\end{document}